\documentclass[fleqn,usenatbib,twocolumn]{mnras}
\pdfoutput=1
\usepackage{newtxtext,newtxmath}

\usepackage[T1]{fontenc}
\usepackage{ae,aecompl}

\usepackage{amsmath}	
\usepackage{amssymb}	
\usepackage[pdftex]{graphicx}
\usepackage{epstopdf}
\usepackage{csquotes}
\usepackage{bm}
\usepackage{threeparttable}
\usepackage{booktabs, makecell}
\usepackage{pdflscape}

\title[SSX-Ray Nebulae in the LMC]{Supersoft X-Ray Nebulae in the Large Magellanic Cloud}

\author[D. A. Farias et al.]{
Diego A. Farias,$^{1,2}$\thanks{E-mail: dnfarias@uc.cl (DNF)}
Alejandro Clocchiatti,$^{1,2}$
Tyrone E. Woods$^{3}$
and Armin Rest$^{4}$
\\
$^{1}$Institute of Astrophysics, Universidad Cat\'olica de Chile, Santiago, Chile \\
$^{2}$Millennium Institute of Astrophysics (MAS), Santiago, Chile\\
$^{3}$National Research Council of Canada, Herzberg Astronomy \& Astrophysics Research Centre,\\ 5071 West Saanich Road, Victoria, BC V9E 2E7, Canada\\
$^{4}$Space Telescope Science Institute, 3700 San Martin Drive, Baltimore, MD 21218, USA
}

\date{Accepted 2020 July 21. Received 2020 July 14; in original form 2020 May 25}

\pubyear{2020}

\begin{document}
\label{firstpage}
\pagerange{\pageref{firstpage}--\pageref{lastpage}}
\maketitle

\begin{abstract}
Supersoft X-rays sources (SSSs) have been proposed as potential Type Ia supernova (SN Ia) progenitors. If such objects are indeed persistently X-ray luminous and embedded in sufficiently dense ISM, they will be surrounded by extended nebular emission. These nebulae should persist even long after a SN Ia explosion, due to the long recombination and cooling times involved. With this in mind, we searched for nebular [\ion{O}{iii}] emission around four SSSs and three SNRs in the Large Magellanic Cloud, using 
the 6.5m Baade telescope at Las Campanas Observatory and the {\sc imacs} camera. We confirm that, out of the four SSS candidates, only CAL 83 can be associated with an [\ion{O}{iii}] nebula. The [\ion{O}{iii}] luminosity for the other objects are constrained to $\lesssim 17$ per cent of that of CAL 83 at 6.8 pc from the central source. Models computed with the photoionization code {\sc cloudy} indicate that either the ISM densities in the environments of CAL 87, RX J0550.0-7151 and RX J0513.9-6951 must be significantly lower than surrounding  CAL 83, or the average X-ray luminosities of these sources over the last $\lesssim$10,000 years must be significantly lower than presently observed, in order to be consistent with the observed luminosity upper limits. For the three SNRs we consider (all with ages $<$1000 years), our [\ion{O}{iii}] flux measurements together with the known surrounding ISM densities strongly constrain the ionizing luminosity of their progenitors in the last several thousand years, independent of the progenitor channel.
\end{abstract}



\section{Introduction}

For many decades the origin of Type Ia supernovae (SNe Ia) has remained elusive.
Early proposals of the explosion
of a white dwarf close to the Chandrasekhar mass
\citep{Hoyle_Fowler_1960} were theoretically confirmed as early as the
1970s \citep{Wheeler_Hansen_1971, Nomoto_Sugimoto_Neo_1976} and the
subsequent progress in both modelling and computing power has allowed for an
ever-improving match between theory and observation \citep{Blondin_Kasen_Ropke_Kirshner_Mandel_2011,Moll_Raskin_Kasen_Woosely_2014,Raskin_Kasen_Moll_Schwab_Woosely_2014,Abigail_Nugent_Kasen_2019}.
Presently, it is understood that a SN Ia is an exploding white
dwarf \citep[see][for a recent review]{Livio_Mazzali_2018}.
The strength of this scenario in explaining  the standardizable energy output of SNe Ia, which led~\citet{Kowal_1968} to highlight their potential as cosmological distance indicators, was exploited to great success with the discovery of Cosmic Acceleration
\citep{Riess_et_al_1998, Perlmutter_et_al_1999}.
\\*

There remains considerable contention, however, in understanding why a white dwarf should explode as a SN Ia. In the classic model \citep{Wheeler_Hansen_1971}, a massive white dwarf accretes matter from an interacting binary companion until triggering an explosion (the Single Degenerate
Scenario, SD); alternatively, \citet{Iben_Tutukov_1984} and \citet{webbink} presented the merging of two
white dwarfs as another pathway to explosion (known as the Double
Degenerate scenario, DD). A number of variations on these two pictures have since emerged \citep[e.g.,][]{Rosswog09,DiStefano11,KS11}, however, we still do not know how a white dwarf may approach the catastrophic instability leading to its explosion as a SN Ia.
\\*

In recent years, both channels have alternatively received observational support and discouragement \citep[for recent reviews of SN Ia progenitor models and observational constraints, see][]{Livio_Mazzali_2018,RuiterReview}. 
Challenging the classical single-degenerate scenario, there have been no observations of surviving companion stars associated with SNe Ia, i.e. near the location of a thermonuclear supernova remnant \citep[hereafter SNR, ][]{edwards,schaefer,RuizLapuenteReview}
and, in general, the lack of hydrogen in SN Ia spectra argue against a main sequence or giant donor consistent with the SD channel.
DD channels largely satisfy these constraints \citep[though may still yield surviving companions, see ][]{Kerzendorf1006}, however the observed global symmetry of SN Ia explosions remains difficult to reconcile with this model \citep{Wang_Wheeler_2008}.
Many approaches have been used to probe the nature of the progenitors of individual known SNe Ia: very early observations of SN 2011fe led to constraints on the size of the exploding star~\citep{nugent2011}, while pre-explosion images set strong upper limits on the luminosity of its possible SD~\citep{graur} and DD progenitor systems~\citep{li2011fe}. The claimed discovery of the companion star of the SN 1572~\citep[Tycho's supernova,][]{ruizlapuente} later proved controversial \citep{kerzendorf}, and subsequent efforts to discover surviving companions have found no other candidates for Tycho or other SN Ia remnants \citep{Kerzendorf1006, RuizLapuenteReview}.
\\*

Searching for evidence of persistent ionizing emission prior to explosion provides a particularly vital test of single degenerate models due to their frequent association with supersoft X-ray sources ~\citep[SSS,][]{kahabka1994, kahabka1997}.
These objects have extremely soft X-ray spectra, much softer than the classical X-ray binaries with neutron star or black hole accretors, and luminosities on the 
order of the Eddington limit \citep{kahabka1997}.
To explain their unique spectra,  
\cite{vanden} were the first to propose that SSSs should be compact objects accreting hydrogen or helium rich material from a companion star at 
the same rate that it burns close
to the surface; this regime is usually called the steady stable-burning-state~\citep[see][for hydrogen and helium accretion, respectively]{sienkiewicz1980, nomoto1982}. 
When hydrogen is accreted (at a specific rate $\dot{M}\sim 10^{-7} M_{\sun} \text{yr}^{-1}$) onto the envelope of the WD ($M_{\text{WD}} \approx 0.7 - 1.2 M_{\sun}$), it burns at a
steady rate and emits significantly in the X-ray band as a blackbody with effective temperatures about $5\times10^5$ K. This rate is approximately that of thermal timescale mass-loss from a moderately evolved donor \citep{nomoto2007}.
The large flux of X-ray radiation resulting from the nuclear-burning at the surface of the WD (a few times $10^{37} \text{ erg s}^{-1}$) can ionize the interstellar medium (ISM) to distances of dozens of parsecs \citep{rappaport}.
Thus, ionized nebulae are expected to be associated with SSSs (also called \enquote*{supersoft nebulae}).
\\*

As opposed to the method of directly detecting their X-ray emission,  
searching for supersoft nebulae has been suggested as a more efficient way to find SSSs,
since the wavelength of typical nebular emission lines are much less affected by ISM absorption than the X-rays \citep{rappaport,kahabka1997}.
For the Large Magellanic Cloud (LMC), in particular, the hydrogen column densities along the line of sight are in the range of $5\times10^{20}$ to $3\times10^{21}\text{ cm}^{-2}$ \citep[see][]{kuuttila2019,SSScolumn},
which implies that the absorption due to the gas principally affects those X-rays with energies less than 0.5 keV.
\citet{remillard} did a search for supersoft nebulae surrounding SSSs in the LMC and found only one, CAL 83, out of nine that they observed.
Two main hypotheses have been proposed to explain this \citep{distefano1995}.
One is that the luminosity of the SSSs varies in time, and that the average is lower than those inferred from the current X-rays observations.
The other is that the surface brightness of the supersoft nebulae is below the thresholds of detection because the ISM around the sources has a very low density.
Both studies tentatively conclude that roughly $10$ per cent of supersoft sources would be associated with ionization nebulae, and that SSSs which are in the steady-burning state are likely to have surrounding nebulae.
\citet{woods2016} developed a statistical argument to conclude that most of the SSS nebulae in the LMC must lie in an ISM much less dense than that of CAL 83. 
Notably, for SN remnants, the expanding shock provides a natural probe of both the surrounding ambient ISM's density and, in some cases, the neutral fraction, eliminating this ambiguity; this has previously been utilized by \cite{woods2017} to 
 exclude most accretion scenarios for the progenitor of Tycho’s supernova. By similar means, \cite{woods2018} and \cite{kuuttila2019} have set constraints on the luminosity of the progenitor of SN 1006, and several SN remnants in the LMC. 
\\*

Our aim in this work is to find nebulae surrounding known SSSs in the LMC and relic nebulae surrounding recent SN remnants.
Our motivation is twofold. On one hand, we want to 
revisit some of the fields studied by \cite{remillard} using more modern instrumentation for observing and more developed software tools for the analysis.
\citet{woods2016} suggested that using 6--8m-class telescopes, and the same observing times of \citet{remillard}, much stricter upper limits on the [\ion{O}{iii}]$\lambda 5007$ {\AA} surface brightness could be established, and that for typical SSS temperatures and luminosities, nebular densities could be constrained down to $n = 0.1 \text{ cm}^{-3}$.
This would mean a factor of four improvement over the upper limits established by \citet{remillard}.
At the same time, we want to find relic nebulae surrounding young Type Ia supernovae remnants.
The very long recombination times at the measured ambient ISM densities for these remnants support the hypothesis that, if a supersoft nebula existed before the explosion, it will remain there for many thousands of years, with the precise timescale depending on the particular optical emission line of interest. Overall, the H-recombination time of any putative photo-ionized nebula is of order $\tau_{\text{rec}} \approx 10^{5} (n_{\text{ISM}}/1 \text{cm}^{-3} )^{-1} \text{ yr}$ \citep{woods2016, woods2017}. For doubly-ionized oxygen, however,  responsible for the strong, collisionally-excited forbidden line [\ion{O}{iii}]$\lambda 5007${\AA} (an important coolant in such nebulae), the recombination timescale is:

\begin{equation}\label{eq:rec}
 \tau_{\text{O}^{2+}\text{rec}} \sim 10^{4} \left ( \frac{n_{\text{ISM}}}{1 \text{cm}^{-3}} \right )^{-1} \text{ yr}.
\end{equation}
\\*

\noindent \citep[see e.g.,][]{osterbrock}. Note here that the cooling time in such a nebula is similarly of order ten thousand years; together these timescales set the observable lifetime for any nebula after the turn-off (or explosion) of its source. For any hot, luminous progenitor scenario, young SNe remnants such as those found in the LMC with ages $\sim500$ yrs \citep{rest} would still be embedded in these associated nebulae long after the explosion.
Finding supersoft nebulae around SSSs and recent Type Ia SN remnants would support the idea that the SSSs are indeed progenitors of SN Ia, and allow one to infer the effective temperatures and luminosities of the progenitors prior to explosion.
An additional strength of observing the LMC is that its distance is well known \citep[$\approx 50$ kpc, see][and references therein]{walkerLMCdistance}. This, together with a precise determination of the surrounding ambient ISM densities \citep[e.g.,][]{kosenko}, ionization fractions \citep[e.g.,][]{Ghavamian2000}, and in some cases light echo spectra of the original supernovae \citep{rest}, allows for a more certain translation between the apparent and intrinsic properties of the source and nebulae and a more precise connection between observations and models.
\\*

We took images of LMC SSSs and recent Type Ia SNRs in a narrow passband centered on the wavelength of the [\ion{O}{iii}]$\lambda 5007${\AA} transition at the redshift of the LMC and the standard $V$ broadband (which contains this line).
The subtraction of these images provides the flux emitted in the line without the contribution from the continuum.
Modeling the measured emission with a photoionization code, we can set constraints on the properties of the nebulae and the SSSs.
In \S 2 we describe the fields observed and in \S 3 we introduce our observations. In \S 4 we discuss the reduction and analysis methods and in \S 5 we present our results and compare them with simple theoretical models.
Finally, in \S 6 and 7 we give a discussion and conclusions of our study.  

\section{Targeted fields}


We selected four SSSs in the LMC, CAL 83, CAL 87, RX J0550.0-7151 and RX J0513.9-6951, as well as three SN remnants, SNR N103B, SNR 0519-69.0 and SNR 0509-67.5.
The four SSSs were observed by \citet{remillard},  who detected an ionization nebula only in CAL 83.
They measured a total [\ion{O}{iii}] luminosity of $1.26\times10^{35}$ erg s$^{-1}$, integrated up to 7.5 pc from the source\footnotemark\footnotetext{Assuming a distance of 55 kpc to the LMC.}.
For the other three sources, they set an upper limit on their bolometric luminosities $L\sim 10^{34.3}$ erg s$^{-1}$, more than a factor ten below the detection in CAL 83, {\it so long as the density of the surrounding ISM is comparable to that in the vicinity of CAL 83.} Higher luminosities, however, remained possible if the surrounding ISM density were lower, and indeed, CAL 83 is now understood to lie in an unusual over-density \citep[see e.g.,][for further discussion]{woods2016}.
\\*

Regarding the SNRs there is an ample consensus that they correspond to Type Ia SN explosions \citep[][]{badenes2009} and no detection of [\ion{O}{iii}] regions around them has been reported.
\citet{kuuttila2019} studied the \ion{He}{ii}$\lambda 4686$ {\AA} surface brightness profiles around these remnants (at $\approx 5$ pc from the source) using IFU observations and put strong upper limits (of order of 10$^{-19}$ erg s$^{-1}$ cm$^{-2}$ arcsec$^{-2}$) in the \ion{He}{ii}  emission line. Their method is similar to that used in this work, but here we focus on the much stronger [\ion{O}{iii}] 5007{\AA} line.
Some additional information on the objects is given below.

\subsection{CAL 83}

CAL 83 was discovered in the original Einstein survey of the LMC \citep{long}. It has been identified as a $17$ mag variable
blue stellar object and is the prototype of the SSS class with orbital period of 1.04 days \citep{hasinger}. 
The images of \citet{remillard} show that the nebula is not symmetric around the source, and the brightness distribution indicate that it is not homogeneous. They do not give an intensity profile but report the integrated luminosities at two different distances (7.5 and 25 pc).
They also report that there is a minimum of emission near the source, where surface brightness falls to $\sim 50$ per cent of the maximum.
\\*

\citet{gruyters} did a spectroscopic study of a subregion of this nebula and, using the ratio of sulphur lines [S II] $\lambda$6716/[S II] $\lambda$6731, estimated a mean electron density of $n\sim 10 \text{ cm}^{-3}$.
This is consistent with that quoted by \citet{remillard} at 7.5 pc\footnotemark[\value{footnote}] from the source. 
The fact that the CAL 83 [\ion{O}{iii}] nebula is not spherically symmetric poses a challenge when modeling it.

\subsection{CAL 87}

CAL 87 was also discovered in the $Einstein$ telescope's first survey \citep{long} and is roughly a factor of four fainter than CAL 83 \citep{hasinger}.
It was identified with a binary star, in this case an eclipsing binary with an eclipse depth of 2 mag, mean $V$ magnitude of $\sim19$ and a period of 10.6 hours \citep{cowley90}.
From observations, it has been suggested that we are seeing only the accretion disk since the WD should be more luminous.
Absorption by the disk would reprocess the soft X-rays from the WD, and explain the missing radiation  \citep{starrfield}.

\subsection{RX J0513.9-6951}

This is the most luminous of the known SSSs in either the Milky Way or the Magellanic Clouds.
Its X-ray variability was discovered during the ROSAT All Sky Survey \citep{schaeidt}.
Its binary companion was identified, with an orbital period of 0.76-days, from optical light curves \citep{hutchings}.
RX J0513.9-6951 displays some unusual characteristics 
such as repeated X-rays outbursts on timescales of years \citep{crampton}.


\subsection{RX J0550.0-7151}
RX J0550.0-7151 was not detected by $Einstein$ but was discovered by \citet{cowley} as a bright and very soft source that lies only $\sim 45$ arc minutes SW of CAL 87, in the same ROSAT frame.
It has been mentioned in fewer references than similar objects in the LMC, which mostly refer to its  symbiotic nature.
\citet{schmidtke1995} first noted that this is a symbiotic system and gave an improved position\footnotemark\footnotetext{RA$_{1950}$: $5^\text{h} 49^\text{m} 46.7^\text{s}$, 
DEC$_{1950}$:$-71\degr49\arcmin38\arcsec$}, which is nearly coincident with a fairly bright red star ($V = 13.53$, $B-V = 1.45$, $U-B = 0.86$).
They, as well as \citet{charles1996}, verified the presence of Balmer emission lines superposed on the spectrum of a cool star, a characteristic feature of symbiotic stars\footnotemark\footnotetext{For a review on the spectra of these objects see \citet{kenyon1984}.}.
The coordinates of this object (transformed to J2000.0 as they appear in the Catalog of Supersoft X-ray sources,  SSSCAT\footnotemark\footnotetext{\href{http://www.mpe.mpg.de/~jcg/sss/ssscat.html}{http://www.mpe.mpg.de/\~jcg/sss/ssscat.html}}), are given in Table~\ref{tab:obs}.
\\*

By 1996, \citet{reinsch1996} found the source in an off state, and there have since been no further X-ray detections. Together with a failed radio detection \citep{fender1998}\footnotemark\footnotetext{Also,they do not obtained emission from CAL 83 and RX J0513.9-6951.}, these later observations have cast some doubt on the nature of this object.
\\*

There is additional confusion regarding RX J0550.0-7151 because its name appears with slightly mismatching figures in different papers and surveys,
likely originating in the difference between the coordinates at epochs 1950.0 and 2000.0.
The name used by \citet{fender1998} was RX J0550-71. SIMBAD associates this name with RX J0550.0-7151\footnotemark\footnotetext{\href{{http://simbad.u-strasbg.fr/simbad/sim-id?Ident=RX+J0550.0-7151&submit=submit+id}}{{http://simbad.u-strasbg.fr/simbad/sim-id?Ident=RX+J0550.0-7151\&submit=submit+id}}}.
\citet{reinsch1996} call it RX J0549.9-7151 and \citet{kahabka2008}, who additionally point out that the SSS does not have an optical counterpart, name it RX J0550.9-7151.
More significantly, while previous works considered this SSS a symbiotic system, 
\citet{schmidtke1999} pointed out that the observation of RX J0050.0-7151 (another mismatch) was blended with RX J0549.8-7150 and that subsequent analysis (H. C. Thomas, 1996, private communication) resolved the two 
sources\footnotemark\footnotetext{This remained a mystery because RX J0550.0-7151 faded below the detection level in the ROSAT All-Sky survey, 1995 data.}.
This field was observed again with Chandra ACIS-S but the source was not detected \citep{orio2007}.

\subsection{SNR N103B}

The young supernova remnant (SNR) N103B is the fourth brightest X-ray remnant in the
LMC \citep{vander}, with a luminosity in the $0.15-4.5$ keV band of $1.5 \times 10^{37}$ erg s$^{-1}$
\citep{hughes}. Its radial extent is about 
3 pc.  From its light echoes, \citet{rest}
estimated the age of N103B to be $860$ yr and confirmed its nature as a SN Ia remnant from the light echo spectra.
The latter has been questioned by \citet[][]{someya}, who studied the ISM surrounding the remnant and found 
that the progenitor consisted of an hydrogen-dominated plasma, suggesting that the explosion should have been a Type II.
Optically, N103B consists of small bright knots
with emission-line spectra typical of SN remnants: [\ion{O}{iii}]$\lambda5007$, 
[\ion{S}{ii}]$\lambda\lambda6716,6731$, H$\alpha$ \citep{hughes}. \citet{snrn103bli} has suggested that
the star closest to the explosion centre may be a candidate companion for this SNR.

\subsection{SNR 0519-69.0}
Its radial size is 3.6 pc \citep{hughes}. Its spectra is Balmer dominated indicating 
conditions of low excitation. Its age, from light echoes,
is $600\pm200$ years \citep{rest}.
\\*

The remnant was studied in X-rays by \citet{hughes},
the SNR is oxygen-poor and iron-rich better consistent with the result of a thermonuclear supernova (SN Ia) \citep{kosenko}.
\citet{li2019} found a star that is a candidate to the companion star in the SD scenario, based on a peculiar radial velocity which deviates more than 2.5$\sigma$ from the mean of the underlying stellar population.

\subsection{SNR 0509-67.5}

This is the youngest and most symmetric of the four LMC SNRs confirmed as SN Ia remnants \citep{edwards}. %
It size is about 3.3 pc \citep{hughes}, and \citet{rest} set an age of $400\pm50$ years from the study of the light echoes.
The spectra is also Balmer dominated. Since no companion has been found near the centre of the SNR \citep{schaefer}, it was tentatively concluded that it is a result of a double degenerate channel, although the SN is spectroscopically classified as an overluminous SN Ia with light echoes~\citep{rest2008}.

\section{Observations}

\begin{figure} 
\includegraphics[width=\columnwidth]{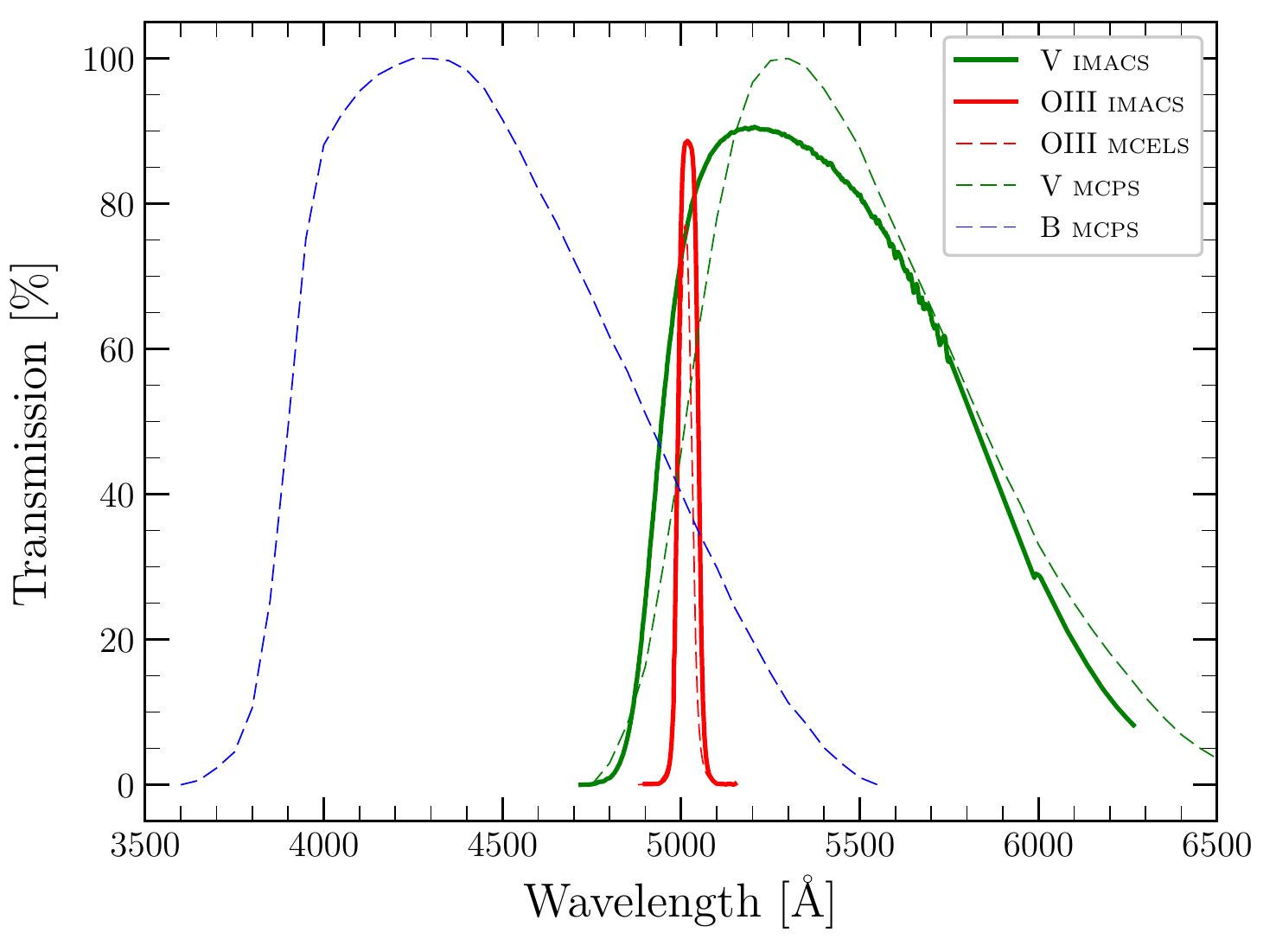}
\caption[{\sc imacs}, {\sc mcps} and {\sc mcels} filters]{{\sc imacs} Bessell $V$ and [\ion{O}{iii}], {\sc mcels} \ion{O}{iii} and {\sc mcps} Johnson $V$ filters.}
\label{fig:filter}
\end{figure}

\begin{figure} 
\includegraphics[width=\columnwidth]{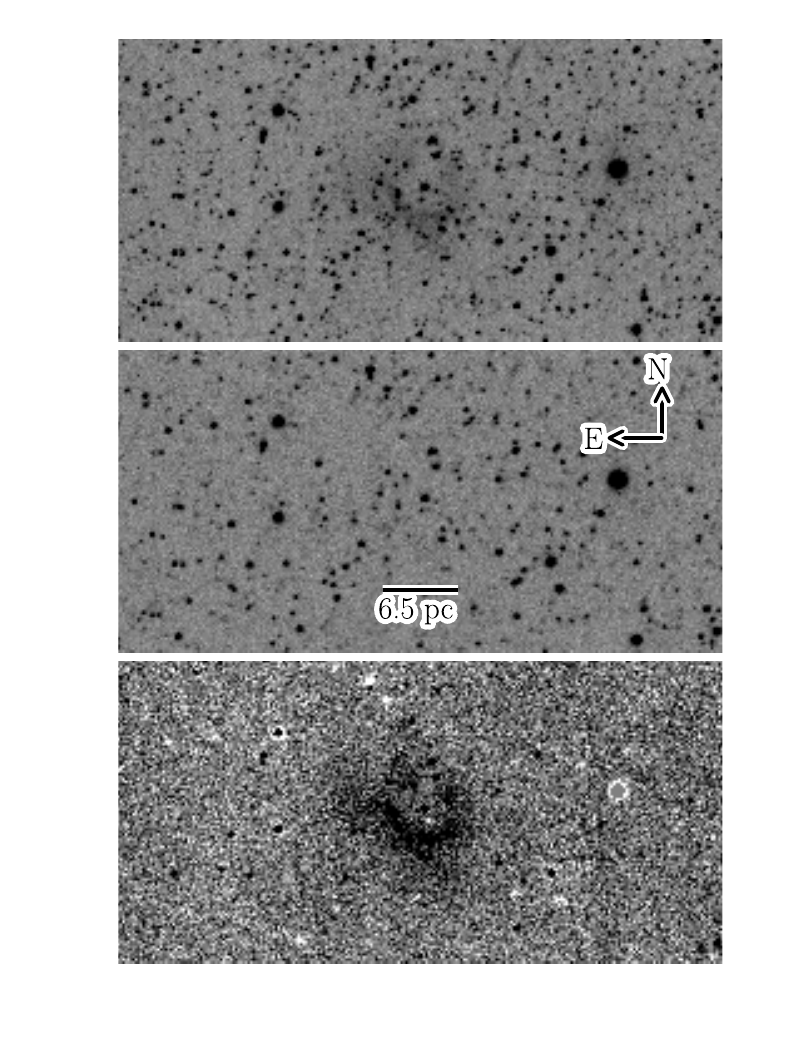}
\caption[\ion{O}{iii}, V and difference images. ]{From up to bottom, {\sc imacs} [\ion{O}{iii}], V and difference images containing the CAL 83 nebula. Dimensions of the images are $\approx 111'' \times 222''$.} 
\label{fig:three}
\end{figure}

Images at two epochs, December 12, 2015 and February 8, 2016, were taken with the Inamori Magellan Areal Camera and Spectrograph 
{\sc imacs}\footnotemark\footnotetext{\href{http://www.lco.cl/telescopes-information/magellan/instruments/imacs/user-manual/the-imacs-user-manual}{{\sc imacs manual}}.} of the Magellan Baade Telescope at LCO, using 
 the all-transmitting (all-spherical optics) f/4.3 long camera (known as the f/4 camera) for direct imaging. The f/4 camera, with eight mosaics 
 of $2k \times 4k$ ({\it MOSAIC3}), observes a $15.4 \times 15.4$ arcmin field, which corresponds to a pixel scale of $0.111$ arcsec per pixel.
Two different filters were used for each observation: the
Johnson-Cousins-Bessell $V$ filter ($5200-7750$ \r{A}) and a custom made narrow band filter centered 
at the observed wavelength of the [\ion{O}{iii}] $5007$\r{A} transition (see Figure~\ref{fig:three}). The 
transmission curves, in comparison with the Magellanic Clouds Emission Line Survey ({\sc mcels}) [\ion{O}{iii}]-filter
and the Magellanic Cloud Photometric System ({\sc mcps}) Johnson $B$ and $V$ filter, can be seen in Figure~\ref{fig:filter}.
To remove the instrumental signatures, bias and flat-field images were taken at each observing epoch. 
A log of the observations is given in Table~\ref{tab:obs}. 
\\
 
\begin{table*}
\caption{Summary of observations.}
\begin{threeparttable}
\begin{tabular}{ l l c c}
\toprule\midrule
 {\sevensize FIELD} & (RA$_{2000}$, DEC$_{2000}$) &  $t_{\text{OIII}}$ (sec)$^{\text{c}}$ & $t_{\text{V}}$ (sec)$^{\text{d}}$ \\
\midrule
 CAL 83$^{\text{a}}$ & (5$^\text{h}$ 43$^\text{m}$ 34.13$^\text{s}$, -68\degr22\arcmin21.9\arcsec)$^{1}$ & $1200$ & $300$ \\
 CAL 87$^{\text{a,b}}$  & (5$^\text{h}$ 46$^\text{m}$ 46.54$^\text{s}$, -71\degr08\arcmin53.9\arcsec)$^{2}$ & $1800,1800$ & $600,300$ \\
 RX J0513.9-6951$^{\text{a}}$ & (5$^\text{h}$ 13$^\text{m}$ 50.8$^\text{s}$, -69\degr51\arcmin47\arcsec)$^{3}$ & $1800$ & $ 600$\\
 RX J0550.0-7151$^{\text{b}}$ & (5$^\text{h}$ 50$^\text{m}$ 0$^\text{s}$, -71\degr52\arcmin9\arcsec)$^{3}$ & $1800$ & $180$ \\
 SNR 0509-67.5$^{\text{a}}$ & (5$^\text{h}$ 09$^\text{m}$ 31$^\text{s}$, 67\degr31\arcmin18\arcsec)$^{4}$ & $1200$ & $ 210$ \\
 SNR 0519-69.0$^{\text{a}}$ & (5$^\text{h}$ 19$^\text{m}$ 35.14$^\text{s}$, -69\degr02\arcmin18\arcsec)$^{5}$ & $1200$ & $300$ \\
 SNR N103B$^{\text{a}}$ & (5$^\text{h}$ 08$^\text{m}$ 59.7$^\text{s}$, -68\degr43\arcmin35.5\arcsec)$^{6}$ & $1200$ & $280$ \\
\bottomrule\addlinespace[1ex]
\end{tabular}
\begin{tablenotes}\footnotesize
\item[] Taken in the $^{\text{a}}$first (12/12/2015) and $^{\text{b}}$second (08/02/2016) epochs.
\item[] Approximated exposure time on $^{\text{c}}$[\ion{O}{iii}] and $^{\text{d}}V$ Bessell filters.
\item[1] \citet{gaia_2018}.
\item[2] \citet{graczyk2011}.
\item[3] From \href{http://www.mpe.mpg.de/~jcg/sss/ssscat.html}{SSSCAT}.
\item[4] \citet{badenes2010}.
\item[5] \citet{fuhrmeister2003}.
\item[6] From \href{https://hea-www.harvard.edu/ChandraSNR/SNRJ0509.0-6843/}{Chandra Supernova Remnant Catalog}.

\end{tablenotes}
\end{threeparttable}
\label{tab:obs}
\end{table*}

Unfortunately, both nights were non-photometric, with low density clouds streaming through the sky. This prevented us from obtaining an absolute 
calibrated photometry and further complicated the analysis.
\\

Note that throughout this work we adopted the most up-to-date distance to the LMC, $D_{\text{LMC}} \approx 5\times 10^4$ pc \citep{walkerLMCdistance}. In order to compare the results with previous studies that used $D_{\text{LMC}} = 5.5\times 10^4 $ pc \citep[e.g., ][]{remillard, 
rappaport,gruyters}, we corrected their projected distances in the LMC (e.g., 7.5 and 25 to 6.8 and 22.7 pc, respectively) and luminosities accordingly.  With a pixel scale of 
$0.111\arcsec$, the projected distance in the LMC of one of our pixels is $0.026$ pc.
%

\section{Reduction and Analysis Methods}
\label{methods}

The basic idea is to isolate the [\ion{O}{iii}] emission by subtracting a conveniently scaled $V$ image from an [\ion{O}{iii}] narrow-filter image.
This should provide the net [\ion{O}{iii}] emission
of any potential ionized nebula surrounding a
SSS or SNR. 
Prior to subtraction, the images must be appropriately processed.

\subsection{Image Processing, alignment and subtraction}

First, the images in both filters were corrected for bias and flat-field using calibration frames taken on the same observing runs and standard {\sc iraf}\footnotemark\footnotetext{\href{http://ast.noao.edu/data/software}{http://ast.noao.edu/data/software}} tasks.
Then, since no WCS information was given in the raw images,
{\sc astrometry.net}\footnotemark\footnotetext{\href{https://github.com/dstndstn/astrometry.net}{https://github.com/dstndstn/astrometry.net}}\citep{astrometrynet} was used to generate precise WCS information from scratch.
The astrometrically aligned images were then fed to the differential photometry code {\sc  photpipe}, a pipeline created by Armin Rest and collaborators \citep[see][and references therein]{arminphotpipe}.
We used {\sc  photpipe} to resample the images of  each field in both filters to a common centre in order to align them in a pixel-wise sense.
With the images aligned we subtracted the $V$ passband from the [\ion{O}{iii}] image using {\sc hotpants}\footnotemark\footnotetext{\href{https://github.com/acbecker/hotpants}{https://github.com/acbecker/hotpants}} \citep{hotpants}, the so called difference images.
The ideal expected output is an image of the field where all the normal stars, together with any thermal continuum emission, have disappeared and the diffuse [\ion{O}{iii}]$\lambda 5007$\r{A} emission of any putative nebula is highlighted.
We found that, as long as the fields are not overly crowded with stars, the subtraction tends to produce reasonable results.

\subsection{Photometric calibration}\label{photocal}

The next step is to find the relation between the counts in the difference image and the received flux in physical units, i.e., the photometric calibration.
We derive it from the definition of count rate cps (typically in units of ADU s$^{-1}$),  of a reference star with spectrum $f_{ \lambda }$ (in units of erg s$^{-1}$ cm$^{-2}$ \AA$^{-1}$), measured under a certain filter $X$, as
\begin{equation}
\text{cps}_{\textit{X}} = \int\limits_{\textit{X}} f_{\lambda} (\lambda) S_{\textit{X}}(\lambda) d\lambda, \label{cpsdef}
\end{equation}
where $S_{\textit{X}}(\lambda) = T_{\textit{X}}(\lambda)\cdot T_{\text{atm,{\it X}}}(\lambda)\cdot R(\lambda)$ is a generalized filter function including the transmission of the atmosphere ($T_{\text{atm,{\it X}}}$, unitless), the optics ($R_{\textit{X}}$, typically in units of ADU erg cm$^{2}$) and the passband $X$ \citep[$T_{\textit{X}}$, unitless,][]{spector2012}.
\\*

In this case, the wide ({\it W}) and narrow ({\it N}) filters are the standard {\sc imacs} Bessell $V$ and our custom made [\ion{O}{iii}], respectively. %
The measured \enquote{cps} values are formed by the contribution of the continuum ($cont$) and the emission line ($line$), as follows
\begin{equation}
   \begin{split}
    \text{cps}_{\textit{N}} &= \text{cps}_{\text{{\it N},cont}} + \text{cps}_{\text{{\it N},line}},\\
    \text{cps}_{\textit{W}} &= \text{cps}_{\text{{\it W},cont}} + \text{cps}_{\text{{\it W},line}}.
    \end{split}
\end{equation}

We follow \citet{bessell2012} and assume that the width of the emission line is much narrower than the effective width of both  [\ion{O}{iii}] ($\Delta_{\text{O}}$) and $V$ ($\Delta_{\textit{V}}$) filters (62 {\AA} and 865 {\AA},  respectively).
%
If so, the flux of the line can be approximated by


\begin{equation}
\begin{split}
   f_{\text{line}} &\cong 
    \frac{\text{cps}_{\text{{\it N},line}}}{S_{\textit{N}}(\text{line})}
  \\
  &\approx \frac{\text{cps}_{\text{{\it N},line}}}
    {T_{\textit{N}}(\lambda_\text{line})T_{\text{atm,{\it N}}}(\lambda_{\text{line}})R(\lambda_{\text{line}})}
    \\
    &\approx \underbrace{\left \{ \frac{\text{cps}_{\textit{N}} - {\text{cps}_{\text{{\it N},cont}}}}{T_{{\it N}}(\lambda_{\text{line}})} \right \} }_{\Delta \text{cps}_{\textit{N}}} 
    \underbrace{\left ( \frac{1}{T_{\text{atm,{\it N}}}(\lambda_{\text{line}})\label{approximation} R(\lambda_{\text{line}}) } \right )}_{\eta_\textit{N}^{-1}}.
\end{split}
\end{equation}

\noindent The first factor in the last expression of Equation~\ref{approximation}, $\Delta \text{cps}_{\textit{N}}$,  can be obtained directly from the image subtraction (we will further assume that $T_{\textit{N}} (\lambda_{\text{line}}) \approx 1$).
The second factor in this equation, $\eta_{\textit{N}}^{-1}$, is the constant that must be found to accomplish the photometric calibration.
Several observations of spectrophotometric stars at different airmass (or at similar atmospheric conditions) are needed to do so.
The main problem with this, ideal, procedure is that the there are no {\sc imacs}  [\ion{O}{iii}] standard stars available so far and, hence, a calibration in the style of \citet[][]{spector2012}
cannot be done.
\\*

On the other hand, the {\sc imacs} Bessell $V$ filter is a standard passband, so its zeropoint can be approximately characterized by, for example, the Johnson filters used in the Magellanic Cloud Photometric Survey ({\sc mcps}\footnotemark\footnotetext{\href{http://svo2.cab.inta-csic.es/svo/theory/fps3/index.php?mode=browse&gname=Misc&gname2=MCPS}{MCPS filters.} }) 
for the catalog of objects in the LMC \citep{zaritsky2004}, with magnitudes placed in the Johnson-Kron-Cousins system \citep{landolt}. Therefore, Johnson {\it U, B} and {\it V} magnitudes are available for the stars in each field.
\\*

Assuming that the (unavailable) {\sc imacs} magnitudes of the stars in each field 
are equal to the {\sc mcps} counterpart, $m_{*}$, for {\sc imacs/mcps $V$} filter and 
that the Vega magnitude are $B = 0$, $V = 0.03$, the Vega calibrated magnitudes can be written as a function of the counts as follows,

\begin{equation}
 m_{*} - m_{\text{VEGA}} = -2.5\log \langle f_{*} \rangle  + 2.5 \log  \langle f_{\text{VEGA}} \rangle, \label{mageq}
\end{equation}
 \citep{bohlinvegamagdef}, where * stands for either $B$ or $V$, $\langle f_{*} \rangle \approx \text{cps}_{\textit{V}}\cdot \Delta_{\textit{V}}^{-1} \cdot \eta_{\textit{V}}^{-1}$ and $f_{\text{VEGA}}$ corresponds to the
spectrum of Vega\footnotemark\footnotetext{\href{ftp://ftp.stsci.edu/cdbs/calspec/}{ftp://ftp.stsci.edu/cdbs/calspec/}}.

Since we do not have useful spectra for any of the stars in the fields, we need to take a further step.
Following the ideas underlying SED reconstruction \citep{Brown2016}, we assume that

\begin{equation}
     \int\limits_{\textit{X}} f_{\lambda} S_{\textit{X}} d\lambda
     \approx f(\lambda_{{\textit{X}}_{\text{eff}}} ) \cdot \Delta_{\textit{X}} \cdot \eta_{\textit{X}} \label{keyapprox}
\end{equation}
where $\lambda_{{X}_{\text{eff}}}$ is the effective wavelength of the filter {\it X}, defined as $\int T_{\textit{X}}f_{*} \lambda d\lambda / \int T_{\textit{X}} f_{*} d\lambda$, and $f(\lambda_{{\textit{X}}_{\text{eff}}})$ is the flux of the star evaluated at this wavelength.
In other words, Equation~\ref{keyapprox} indicates that the mean flux of an object under a filter, characterized by its effective width and wavelength\footnotemark\footnotetext{With the caution that $\lambda_{\text{eff}}$ is not perfect characterization of a filter because it depends on $f_{*}$ \citep{bessell2012}.} can be approximated by the flux of that object evaluated in the effective wavelength. 
\\*

Using Equation~\ref{mageq} for the $B$ and $V$ filters, leads to

\begin{equation}
    \begin{split}
        &f_{*} (\lambda_{\textit{V}_{\text{eff}}} = 5435.5 \text{ \r{A}}) \approx \langle f_{\text{VEGA}} \rangle_{\text{V}} 10^{-0.4 \textit{V}_{*}},\\
        &f_{*} (\lambda_{\textit{B}_{\text{eff}}} = 4392.5 \text{ \AA}) \approx \langle f_{\text{VEGA}} \rangle_{\textit{B}} 10^{-0.4 \textit{B}_{*}},     
    \end{split}
\end{equation}
for $\Delta_{\textit{B}} = 1009 \text{ \r{A}}$ and $\Delta_{\textit{V}} = 864.7 \text{ \r{A}}$.
Now, for each star, $f(\lambda_{\text{O}_{\text{eff}}} = 5019.4 \text{ \r{A}})$ is obtained by a simple interpolation of the fluxes evaluated at \{$V_{\text{eff}}$, $B_{\text{eff}}$\}, as follows

\begin{equation}
    f_{*} (\lambda_{\text{O}_{\text{eff}}}) \approx 
    \left \{ \frac{\delta f_{{*}_{ \{ \textit{B,V} \}_{\text{eff}}} }}
    {\delta_{ \{ \textit{B,V}\}_{\text{eff}}}  } \right \} \times \delta_{ \{ \text{{\it B},O} \}_{\text{eff}} } + f_{*} (\lambda_{{\textit{B}}_{\text{eff}}}), \label{oapprox}
\end{equation}
where $\delta_{\{ \textit{X,Y} \}} =\lambda_{\textit{X}} - \lambda_\textit{Y}$ and $\delta f_{{*}_{\{ \textit{X,Y} \} }} = f_{*} (\lambda_{\textit{X}}) - f_{*}. (\lambda_{\textit{Y}})$.
Therefore, using Equations~\ref{cpsdef},~\ref{keyapprox} and~\ref{oapprox}, ${\eta_{\text{O}}}^{-1}$ is determined as

\begin{equation}
    \eta_{\text{O}}^{-1} = \frac{f_{*}(\lambda_{\text{O}_{\text{eff}}}) \cdot \Delta_{\text{O}}}{\text{cps}_{\text{O}}},
\end{equation}
with $\eta_{\text{O}}^{-1}$ in units of erg s$^{-1}$ cm$^{2}$ ADU.
The time dependency of this equation is ignored since the flux in the difference image is measured in the same scale as the original [\ion{O}{iii}] image.
In order to fit simultaneously all of the stars in the field, we used {\sc scipy} 
Orthogonal Distance Regression (ODR)\footnotemark\footnotetext{\href{https://docs.scipy.org/doc/scipy/reference/odr.html}{https://docs.scipy.org/doc/scipy/reference/odr.html}.}
including the magnitude errors from the catalog.
$\eta_\text{O}^{-1}$ is our photometric zeropoint (Z$_{\text{F}}$).
\\*

The calibration procedure adopted above has been extensively detailed by \citet{waller1990}.
One relevant assumption of the method is that the effective wavelengths of both filters should be close to the wavelength of the line of interest. This is not well satisfied in our case, where the wavelength difference is $\sim$ 400 {\AA}.
As a result, the approximated continuum contribution in Equation \ref{approximation}, $\text{cps}_{ \{ \text{N,cont} \} }$, will not be as well approximated by $\text{cps}_{\text{$V$}}$ as could be, and calibration of the subtraction will be more uncertain than desired.
\citet{remillard} alleviated this problem using a broadband image which was a combination of $B$ and $V$ images, whose effective wavelength lies much closer to the [\ion{O}{iii}] line.
Cloudy weather in our two runs prevented us from collecting the necessary $B$ images.
The preference to calibrate and normalize by [\ion{O}{iii}], although more cumbersome for the lack of calibrated stars in the field, follows the 
historical approach.


\subsection{Flux measurement}\label{fluxCal}

With the counts in the difference images calibrated, the next step is the construction and analysis of the {\sc segmentation} images to determine what is diffuse emission in them.
%
In essence, this is a simultaneous determination of what is background, what is noise and what is emission. The inputs are the image difference provided by {\sc hotpants} and the mask image provided by {\sc photpipe}. 
The mask image removes
pixels with higher noise resulting from the subtraction of bright stars, for example, although some marginal residuals are left.
After experimenting with our data, we found that a useful additional sanity step is to reject pixels outside of the range $-100 \leq \Delta \text{cps} \leq 100$, which are clearly unrelated with real emission.
\\*

We then go through a two step iteration combing the resulting masked difference image with the routine {\sc detect}\_{\sc sources} from the {\sc python} package {\sc photutils detection}\footnotemark\footnotetext{\href{https://photutils.readthedocs.io/en/stable/detection.html}{https://photutils.readthedocs.io/en/stable/detection.html}}, which requires three additional parameters to define an emission region: a signal threshold, a minimum number of neighbouring pixels, {\it npixels}, with signal above this threshold, and a connectivity number.
The latter 
parametrizes the grouping condition, 
set to require either four-connected pixels
touching along their edges, or eight-connected pixels touching along their edges and corners.
After some experimentation, we found that {\it npixels}$= 20$ was a good value for our images because it allows us to reject any residual from subtraction of bright stars left by the mask and previous filtering, since their total areas are composed of $\sim20$ pixels at most (this, in turn, is related with the fact that the typical {\sevensize FWHM} in our images is about $10$ pixels).
A larger value of {\it npixels} may more robustly reject this noise, but is not desirable because this would also reject real emission from the nebula.
Connectivity proved to have a marginal effect, and we decided to set it to the less restrictive value of four.
\\

Setting the threshold for detection requires particular care. %
In principle, this is provided by the statistics of {\sc hotpants} through the parameters {\sevensize MEAN PIX} and {\sevensize STD PIX}.
Basically, {\sevensize MEAN PIX} is the mean of the pixels that were used in the calculation of the kernel to differentiate the images, while {\sevensize STD PIX} is the standard deviation of those pixels.
These values essentially define the background and have a direct impact in what we will later determine to be net emission from the nebula.
%
We determine this in an iterative manner.
First, we do an initial pass through {\sc detect\_sources} using a threshold of one {\sc hotpants} standard deviation (one 
{\sevensize STD PIX}), in order to select pixels that have even a small chance of being emission, calling the remaining pixels as \enquote{background}.
Then, we create a smoothed background image by fitting a linear surface to those pixels (i.e., adjusting functions of the form $Ax + By + C$)\footnotemark\footnotetext{Using code available from: \url{https://gist.github.com/amroamroamro/1db8d69b4b65e8bc66a6}.}. 
Assuming then that this background is sufficiently accurate, we subtract it from the image, recompute the mean value and standard deviation, and redo the {\sc detect\_sources} step, but now using the much more stringent requirement of five standard deviations to reject a pixel.
The means and standard deviations values of the smoothed backgrounds fitted resulted in $-1.5 \lesssim {\rm count} < 1.5$ and  $< 1.5$, respectively, for all cases.
The values of {\sc hotpants} {\sevensize MEAN PIX} and 
{\sevensize STD PIX} for our images are collected in Table~\ref{tab:stats}.

\subsection{Radial Measurement}

The final step is to measure the surface brightness ({\sevensize SB}) profiles in each of the fields.
With the zero points and uncertainties Z$_{\text{F}}$ and $\Delta$Z$_\text{F}$ calculated in Subsection~\ref{photocal},
the transformation to flux is accomplished by multiplying the counts in each pixel \enquote*{i}, $x_{i}$
by Z$_{\text{F}}$,
\begin{equation}
\hat{x}_{i} = \text{Z}_{\text{F}} x_{i}. 
\end{equation}
Error propagation gives the uncertainty of each pixel in flux,
\begin{equation}
 \Delta \hat{x}_{i} = \text{Z}_{\text{F}} \sqrt{ {x_{i}}^{2} \left ( \frac{\Delta \text{Z}_{\text{F}}}{\text{Z}_{\text{F}}} \right )^2 + {\Delta x_{i}}^{2} }.
\end{equation}
\\*

We follow \citet{graurwoods} in measuring [\ion{O}{iii}] surface brightness by doing aperture photometry in annuli centered on the source. We used {\sc python photutils aperture} package\footnotemark\footnotetext{\href{https://photutils.readthedocs.io/en/stable/aperture.html}{https://photutils.readthedocs.io/en/stable/aperture.html}}, and a custom made code\footnotemark\footnotetext{Loosely based on this \href{https://github.com/fbuitrago/SB_profiles/blob/master/calculating_SB.py}{{\sevensize SB} code}.}.
We place the center of the annulus in the object, mask it, 
and obtain the {\sevensize SB} profile as an array of the total counts in each corona.
\\*

Our first measurements of {\sevensize SB} presented us the surprising result that some of the annuli had a negative flux value.
%
We believe this originates in small errors in the image processing that resulted in the subtraction of a background slightly larger than appropriate.
%
Even when our backgrounds were very close to zero, small uncertainties could add up to make the total ``measured'' counts in annuli with little or null emission 
less than zero, a result that is clearly nonphysical.
A simple solution to compensate for this systematic effect is to add a constant to the pixel counts to compensate for this suspected overestimation of the background.
As a rough estimate, we take this constant as the minimum value of the SB profile at any distance between one and twenty five parsecs from the source (after removing the outliers).
Since, in the real world, nothing prevents the annulus with the lowest flux from having a flux larger than zero, our solution implies that, in cases where our background might have been overestimated, our flux measurements will be lower limits to the real flux.

\subsection{Comparison with models}

To gain some physical insight on our measurements of {\sevensize SB} profiles we use version 13.05 of the photoionization code {\sc cloudy}\footnotemark\footnotetext{\href{www.nublado.org}{www.nublado.org}} \citep[][]{cloudy}.
For a given incident spectrum, initial gas density, spherically symmetric density profile and uniform chemical composition, the code computes the radiative transfer through the gas cloud, solving the equations of statistical and thermal equilibrium, thus keeping track of the detailed balance equations for ion population levels.
As a result, the ionization state of the gas and the population of the energy levels are obtained, allowing for a prediction of the emitted spectrum at different distances from the source  \citep{woods2016,woods2017,woods2018}.
\\*

In practice, {\sc cloudy} provides the
emissivity $\epsilon(r)$ of the model nebula, in units of erg cm$^{-3}$ s$^{-1}$.
This emissivity is integrated along the line of sight $l$ in order to obtain the surface brightness profile \citep{woods2016,kuuttila2019},
\begin{equation}
\text{SB}_i (r) = \int_l \frac{\epsilon(r)}{4\pi}dl,
\end{equation}
numerically, we recover equation A1 of \citet{pellegrini},

\begin{equation}
S_i = 2.3504 \cdot 10^{-11} \times \frac{1}{4\pi}\sum_{k} dl_{k}(x,y) \cdot \epsilon_{i}(x,y,z)  \label{eq:pelleq}
\end{equation}
where $S_i$ is now in units of erg s$^{-1}$ cm$^{-2}$ arcsec$^{-2}$, $x,$ and $y$ are such that $r^2 = x^2 + y^2$ (the distance from the source projected in the plane of the sky), and $z$ and $k$ describe the position of cloud elements perpendicular to the plane of the sky.
The constant $2.3504\times 10^{-11}$ is the inverse of the number of square arcseconds that fit into one steradian. 
\\*

The previous equations assume that the nebulae are spherical.
This issue will be relevant when comparing observations and models because, as we will see, 
in the only case where we have a nebula to compare with \citep[CAL 83,][]{remillard}, there is no spherical symmetry.
\\*

We computed models assuming that the central source is a black body, and explored a large range of luminosities, $36 \leq \log L {\, [\rm erg\, s}^{-1}] \leq 38$, and temperatures, $2 \leq T {\, [\rm 10^{5} K]}\leq 7$, corresponding to the lower and upper limits used by \citet{woods2016}.
We assumed that the ISM has an LMC-like chemical composition, with abundances one half the solar values, e.g., [Fe/H]$\sim -0.3$ dex \citep{LMCmetallicity}, and took the neutral hydrogen density ranging from $-1 \leq \log n\leq 1$, in units of $\text{cm}^{-3}$.
\\*

We follow \citet{woods2016} in setting the code-\enquote*{stopping} condition at the distance where the temperature of the nebula falls below $3000$ K.
At these low nebular temperatures, the hydrogen ionization fraction falls below 10\%, meaning that the ionization front is negligible.
A thorough discussion on the usage of {\sc cloudy} models, and the results that are obtained by this procedure,can be found in \cite{woods2018}.

\section{Results}
\label{results}

We will present our results in three stages. We will first analyze visually the difference images to establish if there are systematic defects attributable to the software used, or unusual features in the fields that could bias the radial profiles to be computed later.
Secondly we will describe the radial profiles computed for the different sources and compare them with independent measurements when they are available in the literature.
We will then present our theoretical models for the profiles based on {\sc cloudy} \citep{cloudy} and finalize the section with a discussion.

\subsection{Visual Inspection of Subtractions}

This section is mainly based on the subtractions done with  {\sc hotpants}.
Table~\ref{tab:stats} presents some useful statistics provided directly by the code, such as the {\sevensize MEAN PIX} and {\sevensize STD PIX}, which are inputs of the first masking stage at {\sc segmentation}.
We note that CAL 87 2 and CAL 87 3 are the two different observations of this source, taken at different epochs, which fell on {\sevensize IMACS} chip 2 and 3, respectively.
\\*

We see in Table~\ref{tab:stats} that
SNR 0519-69.0 {\sevensize MEAN PIX} is higher than that of CAL 83, where we already know there is emission.
This could indicate that either the subtraction of this field was bad, with a large number of positive residuals of stars, or that there really is a large area of [\ion{O}{iii}] emission.
The same can be said about SNR N103B and RX J0513.9-6951. All of these fields have large {\sevensize STD PIX} values.
No clear, relevant information regarding the other fields can be extracted from the table.

\begin{table}
\caption{Subtraction statistics for all of the fields:  {\sc hotpants} {\sevensize MEAN PIX} and {\sevensize STD PIX}.}
\begin{threeparttable}
\begin{tabular}{ l c c}
\toprule\midrule
 {\sevensize FIELD} &  {\sevensize MEAN PIX} & {\sevensize STD PIX}\\
\midrule
CAL 83 & 1.711 & 6.611\\
CAL 87 2 & -0.012 & 8.119\\
CAL 87 3 & 0.136 & 4.517\\
RX J0513.9-6951 & 1.822 & 12.046\\
RX J0550.0-7151 & -0.171 & 4.958 \\
SNR 0509-67.5 & 0.425 & 4.388 \\
SNR 0519-69.0 & 3.734 & 11.596 \\
SNR N103B & 3.412 & 13.144 \\
\bottomrule\addlinespace[1ex]
\end{tabular}
\end{threeparttable}
\label{tab:stats}
\end{table}

\subsubsection{SSSs}

\begin{figure*} 
\includegraphics[width=\columnwidth]{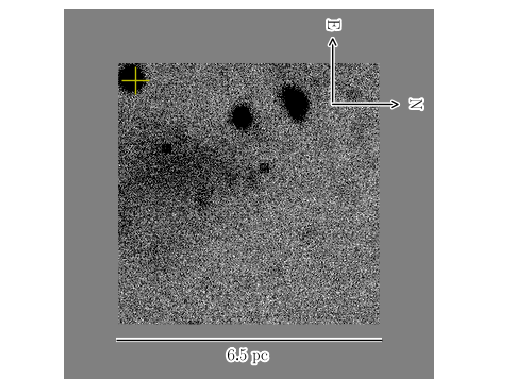}
\includegraphics[width=\columnwidth]{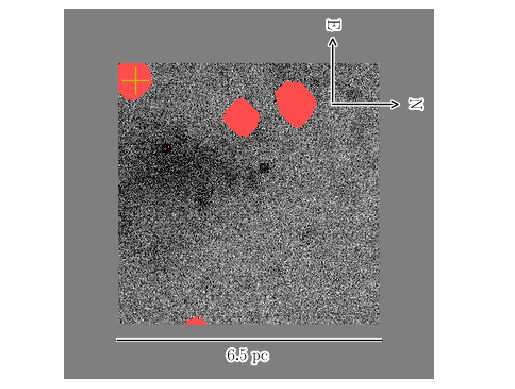}\par\medskip
\caption[Masked and unmasked CAL 83 image] {Difference images of a CAL 83 region near to the source. The left panel shows the image before masking and the right one after applying the masks in the {\sc segmentation} stage (in red). 
The pixel values of the mask are $\approx 0$.
Dimensions of the image are $\approx 6.5\times 6.5$ pc$^{2}$, which makes the region shown approximately the same one studied by \citet{gruyters}.
Diffuse emission is apparent to the eye centered at approximately 3 pc NW from the position of the source (yellow cross).}
\label{fig:gruyters}
\end{figure*}

We will start by analyzing the difference images of CAL 83, in particular the region shown in Figure~\ref{fig:gruyters} which spans a region of a  $\approx 6.5 \times 6.5 \text{ pc}^2$.
This is an area of approximately  $27'' \times 27''$ in the sky, which corresponds closely to the image size discussed by 
\citet{gruyters}.
\\*

These images are well suited to illustrate how the masking of bright objects works.
The typical mask used in the {\sc detect\_sources} task is smaller than required, i.e., some annular shaped residuals of star subtractions remained in the images, therefore it is necessary to enlarge these masks. 
To do this, we used the {\sc python binary dilation} routines.\footnotemark\footnotetext{\href{https://docs.scipy.org/doc/scipy-0.14.0/reference/generated/scipy.ndimage.morphology.binary_dilation.html}{{\sc scipy binary dilation}}}
The dilation process primarily depends on the parameter {\sc iteration}, which is the number of times that the code recursively dilates the mask.
This presents a dilemma: few iterations leave bright rings around the masked objects, but too many iterations (e.g., $\sim 30$), results in large rhomboidal shaped masks at the position of the objects, rejecting from consideration the flux of possible emission pixels.
After some trial and error, we converged on an {\sc iteration} value of $10$ as seen in the right panel of Figure~\ref{fig:gruyters}.
If we compare this image with the left panel, it can be seen that the residuals from stars are completely masked.
For consistency,  we used this value of {\sc iteration} in all of the other fields.
The masked pixel values are set to the mean of the calculated background in {\sc segmentation}, which is always $\approx 0$.\\

In order to judge the quality of the subtraction, we show in the third panel of Figure~\ref{fig:three} the combined [\ion{O}{iii}] (three images of 400 seconds) and the V image (image to be convolved in {\sc photpipe}) covering the location of CAL 83 nebula.\\

\begin{figure} 
\includegraphics[width=\columnwidth]{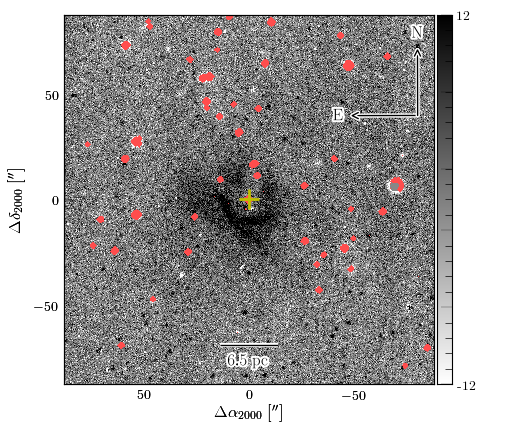}
\caption[{\sc hotpants} CAL 83 subtracted image]{{\sc hotpants} difference image of CAL 83, 
and input of radial measurement, i.e., after masking the objects in {\sc segmentation}. The small yellow cross is the approximated
position of the source. The image is $1600\times 1600 \text{ pixels}^2$, which corresponds to $\approx 178''\times 178''$.} 
\label{fig:CAL83} 
\end{figure}

The full size difference image of CAL 83, applying the masks described above, is seen in Figure~\ref{fig:CAL83}.
This source is not embedded in a crowded field, which facilitates the subtraction and the masking.
The nebula is located surrounding the source, as has been previously characterized by \citet{remillard}.
The diffuse emission is clearly seen both to the NW of the source in Figure~\ref{fig:gruyters} and at the center of the image in Figure~\ref{fig:CAL83}, as a yellow cross.
\\*

\begin{figure} 
\includegraphics[width=\columnwidth]{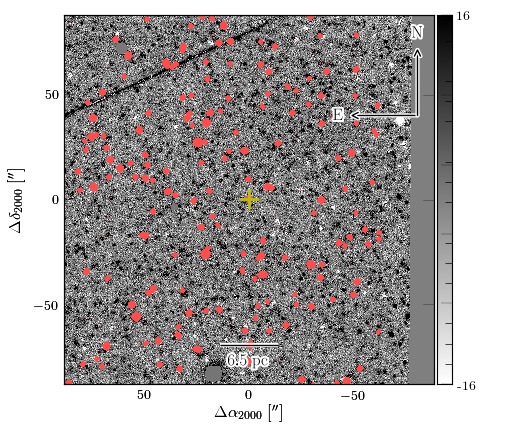}
\caption[{\sc hotpants} CAL 87 chip 2 subtracted image]{Masked {\sc hotpants} output of 
CAL 87 2, ready to be measured. There is no visible
[\ion{O}{iii}] emission as CAL 83 nebula. Black straight line at NE is an artifact in the [\ion{O}{iii}] image.}
\label{fig:CAL872} 
\end{figure}

\begin{figure} 
\includegraphics[width=\columnwidth]{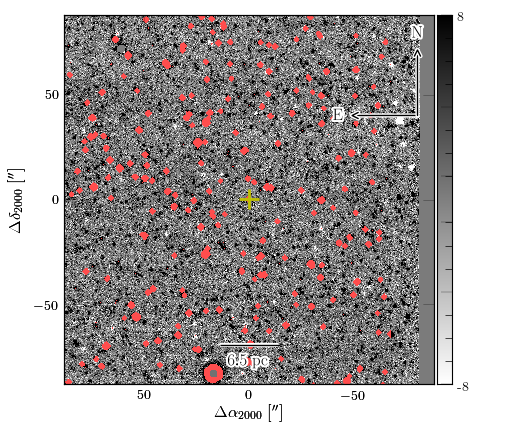}
\caption[{\sc hotpants} CAL 87 chip 3 subtracted image]{Same as Figure~\ref{fig:CAL872} but for CAL 87 chip 3. There is not much
difference with its chip 2 counterpart. We expect that {\sevensize SB/FLUX} profiles have similar shapes for this field, 
but different in scale due to the discrepancy in their corresponding Zs.}
\label{fig:CAL873} 
\end{figure}

CAL 87 presents us with a different scenario. 
We took two images of this source, epoch 1 on chip 2 (Figure~\ref{fig:CAL872}) and epoch 2 on chip 3 (Figure~\ref{fig:CAL873}).
The quality of the subtractions is not as good as that of CAL 83 because the field is more crowded.
Small positive residuals at the positions of stars are abundant.
These residuals are not masked by {\sc segmentation} because they do not match the criteria of being an object
with $20$ or more connected pixels above the threshold.
The comparison of the two independent images of CAL 87 presents us with a good case to understand systematic effects associated with the subtraction technique that will, eventually, appear as differences in the {\sevensize SB/FLUX} profiles.
We note that there is a discrepancy in their zero points, although they are consistent within the uncertainties. 
We will revisit this issue during the analysis of the radial profiles.
Even though the statistics shown in Table~\ref{tab:stats} are computed from the unmasked {\sc hotpants} images, the results  are consistent with the visual analysis.
The {\sevensize MEAN PIX} close to zero in CAL 87 difference images could not have been predicted just by looking at the masked version because of the prevalence of positive residuals of the stars. Therefore, it is clear that the masked objects are negative residuals in order to compensate what we are seeing. 
Nevertheless, the visual inspection indicates that neither of the difference images shows any kind of extended emission consistent with a nebula.\\

\begin{figure} 

\includegraphics[width=\columnwidth]{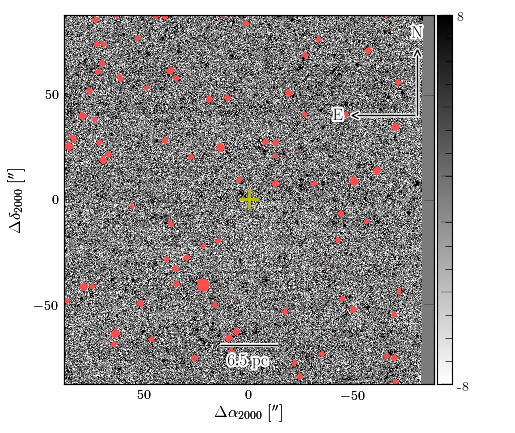}
\caption[{\sc hotpants} RX J0550.0-7151 subtracted image]{{\sc hotpants} RX J0550.0-7151 field image, after being mask
in {\sc segmentation}; visually, this subtraction looks cleaner than CAL 87, as there are not as many residuals, positive or negative,  
of stars, while the masked objects are few, such in the case of CAL 83.}
\label{fig:RXJ0550} 
\end{figure}

The difference image of RX J0550.0-7151, which was taken on epoch 2, is shown in Figure~\ref{fig:RXJ0550}.
It can be seen that it looks cleaner than those of CAL 87, 
with fewer masked objects.
Again, no sign of extended emission surrounding the source, marked by a yellow cross, is seen.
The {\sevensize MEAN PIX} value is negative, but also the closest to zero of all the fields but CAL 87.
This is consistent with the absence of positive residuals from stars after masking.\\

\begin{figure} 
\includegraphics[width=\columnwidth]{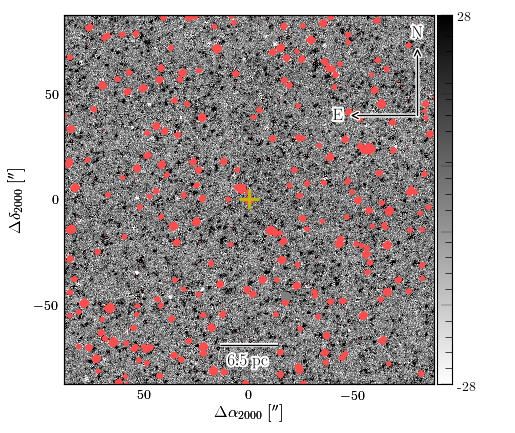}
\caption[{\sc hotpants} RX J0513.9-76951 subtracted image]{Same as previous difference images, for RX J0513.9-6951 field. It is remarkable the
similitude between CAL 87 (Figures~\ref{fig:CAL872} and~\ref{fig:CAL873}) and this image.}
\label{fig:RXJ0513} 
\end{figure}

The last of the SSSs is RX J0513.9-6951 in Figure~\ref{fig:RXJ0513}.
The difference image is similar to that of CAL 87, 
with many residuals of stars and several masked stars.
Although, in contrast to CAL 87, the 
large {\sevensize MEAN PIX} value shown
in Table~\ref{tab:stats} tell us that many masked stars were positive residuals.
\\*

Overall, the visual inspection of the difference images corresponding to SSSs leads us to the same conclusion of \citet{remillard}: There is no nebular [\ion{O}{iii}] emission detected surrounding
them, with the only exception being CAL 83.

\subsubsection{SNRs}\label{SNRs}

\begin{figure} 
\includegraphics[width=\columnwidth]{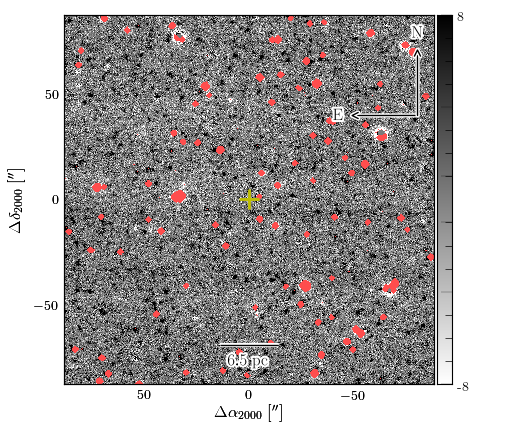}
\caption[{\sc hotpants} SNR 0509-67.5 subtracted image]{Difference image of SNR 0509-67.5; the resemblance with the (now prototype) CAL 87
counterpart is evident, which leads us to conclude that, at least visually, there is no [\ion{O}{iii}] emission from this field.}
\label{fig:SNR0509} 
\end{figure}

The three SNRs in our sample show similar results.
SNR 0509-67.5, shown in Figure~\ref{fig:SNR0509}, appears to be similar to
CAL 87, both in the number of masked stars and the number and shape of the poorly subtracted stars.
This fact, plus the {\sevensize MEAN PIX} value registered in Table~\ref{tab:stats}, suggests no emission at all.

\begin{figure} 
\includegraphics[width=\columnwidth]{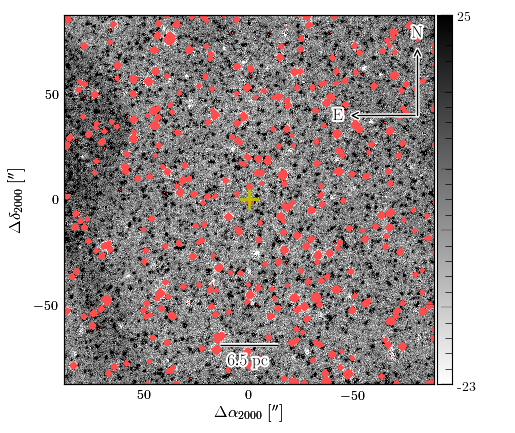}
\caption[{\sc hotpants} SNR 0519-69.0 subtracted image]{Difference image of SNR 0519-69.0 field. This is by far 
one of the most contaminated. See the details in Subsection \ref{SNRs}. 
}
\label{fig:SNR0519} 
\end{figure}

SNR 0519-69.0, shown in Figure~\ref{fig:SNR0519}, is interesting.
The large {\sevensize MEAN PIX} value in Table~\ref{tab:stats} suggests, as in the case of CAL 83, that either there is some emission or the subtraction was very poor. 
Inspection of the difference image suggests that the latter is the most probable case. There is a large emission-like region in the left part of the 
image. What appears to be a circular object in the upper section is an instrumental artifact, probably related with a reflection in the [\ion{O}{iii}] 
filter that is not compensated by flat-fielding (it is seen in other chips of the CCD array in some of the other fields).
These  structures increase the value of {\sevensize MEAN PIX} accordingly.
As for the rest, there are only masked pixels and residual stars surrounding the source, which also contribute to increasing the mean pixel value.\\

\begin{figure} 
\includegraphics[width=\columnwidth]{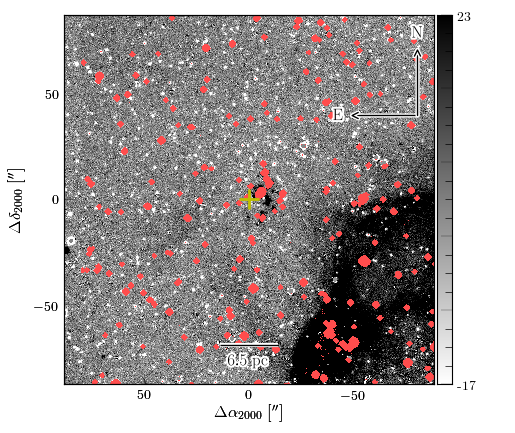}
\caption[{\sc hotpants} SNR N103B subtracted image]{Difference image of SNR N103B.
The complexity of the region is apparent. The superbubble is easily seen at SW. More details in 
Subsection~\ref{SNRs}.%
}
\label{fig:SNRN103B} 
\end{figure}

The last SNR of the set is SNR N103B, also known as SNR 0509-68.7, which is shown in Figure~\ref{fig:SNRN103B}.
This case shows a real detection of [\ion{O}{iii}] emission.
First, the small-scale emission very close to center of the SNR is masked by the code, but a few small knots can be seen outside of the mask, near the location of the source marked by a yellow cross. These structures are well inside the outgoing shock of the remnant, and surrounded by many negative residuals from star subtractions.
It is known that SNR N103B is a very complex structure \citep[see][to appreciate all the physical structures of this SNR]{snrn103bli}, but overall the local emission is overshadowed by that from an extended object seen towards the SW.
This is a superbubble surrounding the stellar cluster NGC 1850\footnotemark\footnotetext{\href{http://simbad.u-strasbg.fr/simbad/sim-id?Ident=NGC+1850}{http://simbad.u-strasbg.fr/simbad/sim-id?Ident=NGC+1850}}. 
Its contribution will certainly be noticed at $\approx 50''\sim 12$ pc from the source.

\subsection{Emission profiles}
\label{se:empirical_profiles}
We present here our measurements of [\ion{O}{iii}] emission radial profiles to be compared later with theoretical models.
We estimate the energy summing total counts in annuli and using the conversion factor Z$_{\text{F}}$ calculated as explained in Section~\ref{photocal}.
The values of Z$_{\text{F}}$ and its uncertainty $\Delta$Z$_{\text{F}}$ for each source are given in Table~\ref{tab:zeros}.
When measuring radially over the nebula, or suspected nebular region, we propagate the uncertainty $\Delta$Z$_{\text{F}}$.
In addition to the counts, it is necessary for constructing radial profiles to determine the distance from the source to each annulus.
The position of the sources are given in columns two and three in Table~\ref{tab:obs}.\\

\begin{table}
\caption{Calculated Z$_{\text{F}}$ and $\Delta$Z$_{\text{F}}$ in units of erg s$^{-1}$ cm$^{-2}$ {\sevensize COUNTS}$^{-1}$.}
\begin{threeparttable}
\begin{tabular}{ l c c c}
\toprule\midrule
{\sevensize FIELD} & Z$_{\text{F}} \times 10^{19}$ & $\Delta$Z$_{\text{F}} \times 10^{19}$ & {\sevensize NUMBER OF STARS}\\
\midrule
CAL 83 & 2.022 & 0.458 & 773 \\
CAL 87 2 & 1.017 & 0.246 & 1315\\
CAL 87 3 & 1.515 & 0.353 & 1367\\
RX J0513.9-6951 & 0.747 & 0.180 & 1062\\
RX J0550.0-7151 & 1.412 & 0.376 & 452\\
SNR 0509-67.5 & 1.407 & 0.332 & 612 \\
SNR 0519-69.0 & 1.213 & 0.273 & 1414\\
SNR N103B & 2.573 & 0.520 & 1684 \\
\bottomrule\addlinespace[1ex]
\end{tabular}
\end{threeparttable}
\label{tab:zeros}
\end{table}

In the absence of a better option, we choose as pedestal the lowest mean value of the annuli.
Table~\ref{tab:annuli} provides the statistics for the annuli in all fields. {\sevensize MIN} is the value of the pedestal subtracted.
Most of the pedestals are consistent with zero, the exceptions being RX J0513.9-6951 and SNR N103B.
The latter is by far the largest departure from zero of the list, but this is not surprising 
because its difference image indicates a problem with the subtraction, with many negative residuals on the stars surrounding the source,
while the large
emission region appears in the outskirts of the image.
The pixels in this distant region add up to a larger positive value that compensates the negative 
isolated residuals, and therefore the field has a
positive {\sevensize MEAN}.
Visual inspection clearly indicates the location of the 
{\sevensize MIN} annulus and strongly suggests why it is negative.
But the difficulty of masking the small residuals leads us to prefer using the pedestal instead of excising the individual residuals for a cleaner background subtraction.\\

\begin{table}
\caption{Corona statistics for all of the fields: minimum {\sevensize MIN}, maximum {\sevensize MAX}, 
mean {\sevensize MEAN} and standard deviation {\sevensize STD}.}
\begin{threeparttable}
\begin{tabular}{ l c c c c}
\toprule\midrule
 {\sevensize FIELD}    & \scriptsize{MIN}  & {\sevensize MAX}  & {\sevensize MEAN} & {\sevensize STD}\\
\midrule
CAL 83 & -0.252 & 10.199 & 1.751 & 2.497\\
CAL 87 2 & -0.080 & 2.554 & 1.201 & 0.607\\\
CAL 87 3 & -0.280 & 1.445 & 0.688 & 0.336\\
RX J0513.9-6951 & 0.838 & 4.746 & 2.366 & 0.733\\
RX J0550.0-7151 & -0.294 & 0.649 & 0.068 & 0.140\\
SNR 0509-67.5 & -0.305 & 0.678 & 0.231 & 0.209\\
SNR 0519-69.0 & 0.041 & 5.337 & 2.422 & 1.476\\
SNR N103B & -2.504 & 6.108 & 2.317 & 2.550\\
\bottomrule\addlinespace[1ex]
\end{tabular}
\end{threeparttable}
\label{tab:annuli}
\end{table}

Figure~\ref{fig:SSScombination1} displays
the final {\sevensize SB} profiles for
the SSSs in our fields, while Figure~\ref{fig:SNRcombination2} does for the SNRs, 
alongside some {\sc cloudy} models (see subsection \ref{subcloudy}). 
For the fields with clear visual evidence of 
[\ion{O}{iii}] emission, the conclusions of our initial inspections in the previous section are confirmed: CAL 83, with a peak at $\approx 3$ pc and a minimum at $\leq 1$ pc from the source (given its doughnut shape), and the superbubble in the field of SNR N103B which contributes effectively from  $\approx 10$ pc.
The rest of the sources do not show clear signs of emission.
SNR 0519-69.0, the curious case that we named the worst subtraction of the list, shows an increment in the {\sevensize SB} profile starting from $\approx 20$ pc,
which we associate with the instrumental artifacts discussed in the visual analysis.
In any case, with the exception of CAL 83, none of the
SNRs or SSSs show signs of an emission nebula around the source.
This is consistent with the conclusions of the earlier work of \cite{remillard} for the objects in common (i.e. the four SSSs).
\\*

Our results on the CAL 83 SSS nebula can be compared
with those of the two previous studies examining it.
\citet{remillard} studied CAL 83, CAL 87, RX J0550.0-7151 and RX J0513.9-6951.
For the former, they defined an inner region for the nebula extending up to 6.8 pc\footnotemark\footnotetext{Originally, 7.5 pc in their work.} from the source and indicated that, within it, the emission reaches a maximum at $\sim 3$ pc.
They also commented that the region closer to 1 pc from the source represents a local minimum.
Both of these observations are verified in this work,
as seen in Figures~\ref{fig:CAL83} and~\ref{fig:SSScombination1}.
It is necessary to say that one of the observations made by
\citet{remillard} is not found in our profiles: they stated that, 
even if there is a local minimum close to 1 pc, it should be 
no less than 50 per cent of the intensity obtained at 3 pc; the reason for such a discrepancy is unclear but bears future study, but it is sure that 
masking and setting the values with the global mean must be taken into account. 
\citet{remillard} also stated that the inner region contributes about 50 per cent of the total flux measured
up to 22.7 pc.
This amounts to $\sim 4\times 10^{-13}$ and $8.2 \times 10^{-13}$ erg s$^{-1}$ cm$^{2} $ respectively,
neglecting the correction factor of 1.4 to account for Milky Way extinction (though they included this later when putting upper limits on the luminosity of the sources).\\

Finally, \cite{remillard} set an upper limit to the luminosity 
of all of the other sources at $6.8$ pc, the count rate of those sources is about a factor ten smaller than that of the CAL 83 inner nebula.
 They estimated that the total [\ion{O}{iii}] luminosity of CAL 83 at 22.7 pc was 
$3.4 \times 10^{35}$ erg s$^{-1}$, corresponding to an upper limit of $\approx 10^{34.22}$  at 6.8 pc for the rest of the sources. \\

Our own measurements for CAL 83 are 
$0.91\times 10^{35}$ and 
$1.73\times10^{35}$ erg s$^{-1}$ at 6.8 and 22.7 pc, respectively, using the same factor to correct for foreground extinction.
Therefore, the upper limit of Remillard is about $17\%$ of our measurement of CAL 83 at 6.8 pc (see  Table~\ref{tab:final})
.   
\\*

We can also compare with the results of \citet{gruyters}
who provide the only other published study of CAL 83 nebula.
They used an IFU spectrum to study the flux emitted in about one quarter of the inner nebula, approximately the region displayed in Figure~\ref{fig:gruyters}.
A main conclusion of their work is that the total dereddened (reddened) [\ion{O}{iii}] flux in their quadrant is
$\approx 9.7\times10^{-14}$ ($\approx6.5\times10^{-14}$) erg s$^{-1}$ cm$^{-2}$. 
They state that the quadrant has dimensions of $25.5\arcsec\times 25.5\arcsec$, 
or equal to $7.5\times7.5$ pc$^{2}$ at a distance to LMC of 55 kpc\footnote{There seems to be a mistake here. At 55 kpc
$25.5\arcsec\times25.5\arcsec$
corresponds to $\approx 6.8 \times 6.8$ pc$^{2}$, 
while their full FoV is $27\arcsec\times 27\arcsec$ ($7.2 \times 7.2 \text{ pc}^2$).}.
Our measurement was done for an 
aperture of $27\arcsec\times27\arcsec$ (full FOV of their work),  $6.5\times6.5 \text{ pc}^{2}$ given the distance correction, centered to approximately match that of \citet{gruyters}, since they do not provide the centre for the aligned
images.
\\*

Our result for this aperture is about $\approx 3.3\times 10^{-14} \text{ erg  s}^{-1}$ cm$^{-2}$ for masked data, a value smaller than the reddened value of \citet{gruyters} (about 51\% of their flux).
If we compute the profile without applying any mask to the data (which considers several positive residuals), other than the one corresponding to {\sc hotpants}, and do not do background subtraction, 
the change of the total flux is about 37 \% larger ($\approx$ 88 \% the reddened value of Gruyters).
Our measurements, hence, recover just a fraction of the flux of either \citet{remillard} at 6.8 pc or \citet{gruyters} closer to the source and appears to produce systematically low results.

\subsection{Theoretical Modelling}\label{subcloudy}

\begin{figure} 
\includegraphics[width=\columnwidth]{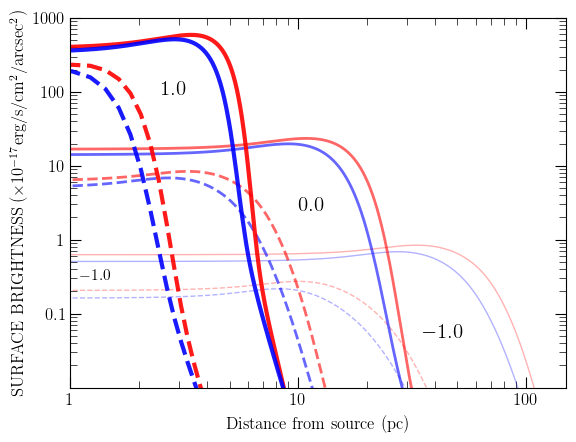}
\caption[{\sc cloudy} {\sevensize SB} profiles]{{\sc cloudy} {\sevensize SB} profiles for $\hat{n}\equiv\log_{10} n = \{ -1,0,1\}$ (ISM density), 
$\hat{L} \equiv \log_{10} L = \{ 37,38\}$ (luminosity of the source in 
dashed and solid lines, respectively) and $T_5 = T\times 10^{-5} = \{ 4,5\}$ (temperature of the source in red and blue lines, respectively), with ranges of
input parameters according to typical densities of ISM, and $T,L$ of a SSS. These models were computed assuming LMC abundances (one half solar metallicity).
}
\label{fig:SBcloudy} 
\end{figure}

We are now ready to compare the
{\sevensize SB/FLUX} radial profiles with {\sc cloudy} theoretical models.
The critical input parameters which we will vary throughout are the density of the ISM ($n$), and the temperature ($T$) and total luminosity ($L$) of the central source.
The chemical composition, the other relevant parameter for {\sc cloudy} models in this configuration, was fixed at one half solar abundance (typical of the LMC).
Figure~\ref{fig:SBcloudy} illustrates the results of a study combining three ISM densities that span those expected for the LMC, with two radiation temperatures and two luminosities for the central source that span the characteristic range observed for SSSs.
Curves in the figure are labeled by the values of $\hat{n} = \log{n}$, with $n$ in units of cm$^{-3}$, line-style coded according to the values of $\hat{L} = \log{L}$, with $L$ in units of erg s$^{-1}$, and color coded according to the values of $T_5 = T \times 10^{-5}$, the radiation temperature of the source, with $T$ in Kelvin.\\

\begin{figure} 
\includegraphics[width=\columnwidth]{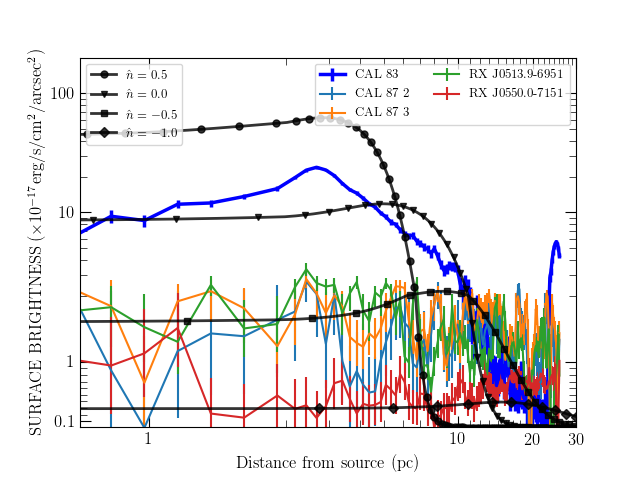}
\caption[{\sc cloudy} and empirical {\sevensize SB} profiles of the SSSs in our sample, for variable $n$, fixed $T_5=5$, $\hat{L}=37.5$]{Empirical surface brightness profiles of all SSSs and {\sc cloudy} 
models for $\hat{n}=\{-1.0,-0.5,0.0,0.5\}$, $\hat{L}=37.5$  and $T_5=5$. The fact that only one profile, CAL 83, lies for sure above $n=1 \text{ cm}^{-3}$ at mostly any distance from the central source,
suggest that in normal conditions, 
there is no nebular emission surrounding the other sources. Otherwise, 
it would required a really low ISM density to explain these {\sevensize SB}
profiles in the figure, or a bad subtraction and/or flux calibration.} 
\label{fig:SSScombination1} 
\end{figure}

\begin{figure} 
\includegraphics[width=\columnwidth]{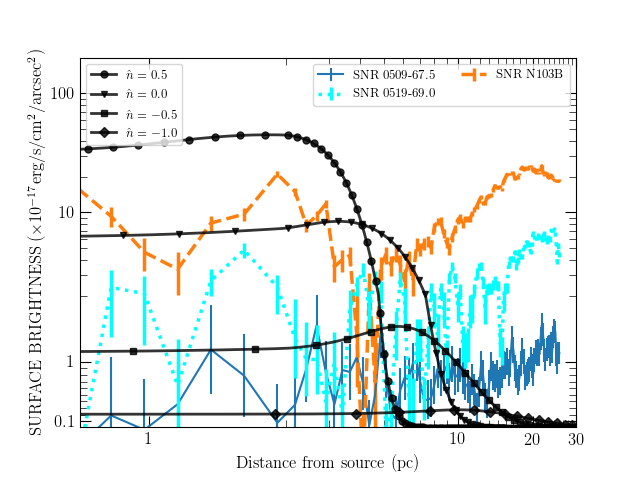}
\caption[{\sc cloudy} and empirical {\sevensize SB} profiles for fixed $\hat{L}=37$]{
Same as Figure~\ref{fig:SSScombination1}, but with the empirical models of our SNRs and models calculated for fixed values $\hat{L}=37$ and $T_5=4$ . In this case, we have more possibilities to explain the lack of emission for SNR 0509-67.5 and SNR 0519-69.0: 1) the ISM is too low, in concordance to the SSS case and 2) the time spent between the SSS phase and the SN explosion is comparable to the recombination time of the ISM.
}
\label{fig:SNRcombination2} 
\end{figure}

Such models may be readily compared with our observations. We have plotted in Figure~\ref{fig:SSScombination1} 
all the empirical {\sevensize SB} profiles of the SSSs obtained in Section~\ref{se:empirical_profiles}, together with illustrative {\sc cloudy} models computed for $\hat{L}= 37.5$ and $T_5=5$.
All of the SSS {\sevensize SB} profiles but CAL 83 are consistent with very low values of $n$, firmly excluding $n \gtrsim 1.0$ cm$^{-3}$, and with many cases being consistent with the range
$ 0.1 \lesssim n \lesssim 0.3$ cm$^{-3}$.
Comparing with the Figure~\ref{fig:SNRcombination2}, the {\sevensize SB} profiles of all of the SNRs, we see that the only field with prominent [\ion{O}{iii}] emission like CAL 83 is the particular case of SNR N103B. This is because at small distances from N103B, some emission from its unmasked knots remains while, at larger distances, it is difficult to separate 
any emission due to an hypothetical SSS fossil nebula from the contribution of the superbubble surrounding NGC 1850. Notice that the {\sc cloudy} models of Figure~\ref{fig:SNRcombination2} were calculated for the fixed values $\hat{L} = 37$ and $T_5 = 4$: as the most important parameter is the density, the conclusion of the SSS case remains, that for SNR 0509-67.5 and SNR 0519-69.0 the models that remain consistent are only those with $n \lesssim 1 \text{ cm}^{-3}$.  
\\*

Taking the offset in flux measurement that we found in Section~\ref{se:empirical_profiles}, between $\approx 1.2$ and $\approx 2$, at face value, we could increase all the profiles by a factor between 20 percent and 100 per cent in order to account for possible error in our calibration.
We would find, then, that almost all the fields would still fall below $n=1 \text{ cm}^{-3}$. %
In the case of CAL 83, applying a factor two to the profile only just brings the observed profile into agreement with the $\hat{n} = 0.5$ model (that is, $n = 3.16 \text{ cm}^{-3}$ at 6.8 pc).
%
This is marginally consistent with the lower bound of the interval $4-10$ cm$^{-3}$ quoted by \citet{remillard} and clearly smaller than the 
$\approx 10 \text{ cm}^{-3}$ given by \citet{gruyters} (note however that the ISM in the immediate vicinity of CAL 83 is understood to be inhomogeneous, see \cite{remillard} and \cite{gruyters} for further discussion).


As we can see in Figure~\ref{fig:SBcloudy}, some degeneracy exists between the $n$, $L$, and $T$ input parameters.
A higher luminosity model will be characterized by slightly more [\ion{O}{iii}] emission 
nearby the source and a generally more extended ionized region.
In such a case, the CAL 83 profile is going to better represented by a model with a lower density than the previously obtained ($1 \lesssim n \lesssim 3 \text{ cm}^{-3}$), although this would be inconsistent with e.g., H$\alpha$ measurements \citep{remillard}, and may be excluded.
The road to a source with higher luminosity seems to be closed, but that of lower luminosity sources appears to be possible.
Decreasing the luminosity of the source and at the same time its radiation temperature, so as not to decrease the size of the ionized region, appears as an interesting option.
With respect to the {\sc cloudy} for $\hat{L}=37$, $T_5=4$ in Figure~\ref{fig:SNRcombination2}, even if the {\sevensize SB} profiles of no-emission fields are better 
constrained by $n\approx 0.3 \text{ cm}^{-3}$ {\sc cloudy} model (i.e. a model than behaved as an upper limit before provides now a better match), the shape of the densest profiles provide a poorer match to the CAL 83 {\sevensize SB} profile.
In Figure~\ref{fig:SSScombination1}, 
the profile $n=1 \text{ cm}^{-3}$ provided a good qualitative fit to the observed {\sevensize SB} profile, and approximately matched the extension of the nebulae.
We can therefore conclude that $\hat{L} \approx 37.5$ is a better value for the time-averaged luminosity of a typical source, at least for a nebula like CAL 83. 
\\\\

\begin{figure} 
\includegraphics[width=\columnwidth]{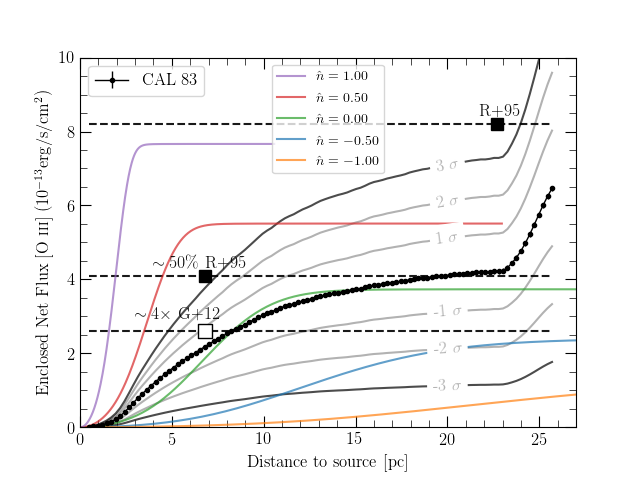}
\caption[CAL 83 {\sevensize FLUX} profile and {\sc cloudy FLUX} models for variable $n$]{CAL 83 {\sevensize FLUX} profile compared with the measurements of \citet{remillard} \& \citet{gruyters} flux values at 6.8 and 22.7 pc from the source, and {\sc cloudy FLUX} profiles, 
for $\hat{n}=\{-1.0,-0.5,0.0,0.5,1.0\}$, $T_5=5$ and $\hat{L}=37.5$.
Our flux measurements are below those of \citet{remillard} by $\sim 45$ per cent at 6.8 pc and $\sim 50$ per cent at 22.7 pc.
The mismatch, though, is within  $3\sigma$ of our uncertainty in the calibration for this field (c.f.~\ref{tab:zeros}).
At 6.8 pc our value is about the same as that of \citet{gruyters} (under the assumption of spherically symmetric emission).}
\label{fig:fluxcal83} 
\end{figure}

Figure~\ref{fig:fluxcal83} presents a study of the {\sevensize FLUX} enclosed in apertures of increasing size for CAL 83 compared with both {\sc cloudy} models and
the values $8.2\times 10^{-13} \text{ erg s}^{-1} \text{cm}^{-2}$ and  $\approx 4\times 10^{-13} \text{ erg s}^{-1}\text{ cm}^2$, which are the measurements provided by \citet{remillard} for the flux enclosed at 22.7 and 6.8 pc, respectively.
The {\sc cloudy} models of this figure were computed for a temperature $T_5=5$, luminosity $\hat{L}=37.5$, and five different densities $\hat{n}=\{-1.0,-0.5,0.0,0.5,1.0\}$.
It is clear in the figure that $n=1 \text{ cm}^{-3}$ provides the profile that better matches the observations, but that a value slightly larger would provide an even better match.
On the other hand, the differences in shape between the empirical and model profiles probably indicate that the nebula does not have a constant density (see discussion above).
\\*

Figure~\ref{fig:fluxcal83} also sheds some light onto the mismatch reported in Section~\ref{se:empirical_profiles} between the inner estimate of \citet{remillard}, consistent with that of \citet{gruyters}, and our own measurement.
They reach the 50 per cent value given by the lower horizontal line at 6.8 pc and we do at $\sim20$ pc.
At 6.8 pc we are about 45 per cent below their estimate.
Our measurement, however, is below theirs by 50 per cent at 22.7 pc.
We are aware that the more precise flux in \citet{remillard} is enclosed within the 22.7 pc aperture than the one within the 6.8 pc aperture\footnote{They write \enquote*{The brighter inner nebula of CAL 83, which we measure by integrating the surface brightness out to a radius of 7.5 pc (6.8 pc) from the central object, contributes $\sim$50 per cent  of the total flux.}}
(Figure~\ref{fig:fluxcal83}), we take the more conservative value at 6.8 pc, meaning that our calibration is underestimated by $\sim$45 per cent.
The differences, however, are within $3\sigma$ of the uncertainty in our zero point for this field at 6.8 pc (c.f.~\ref{tab:zeros}), as the $\sigma$ deviations plotted in the Figure indicate.
\citet{gruyters} studied one quadrant of the CAL 83 nebula and provide a total [\ion{O}{iii}] flux of $6.52 \times 10^{-14}$ erg s$^{-1}$, within 6.8 pc of the central source.
We have plotted this measurement as well in Figure~\ref{fig:fluxcal83}, assuming that the emission is spherically symmetric (although this is clearly only a crude approximation for CAL 83). 
\\*

Our measurements, then, are consistent within the uncertainties with those of previous work. Since our relative flux calibration is more precise than our absolute flux calibration, the different discrepancies with \citet{remillard} at different distances from the source indicate something more complex than just a difference of $\sim 50$ per cent in the zeropoints and the large masked zones.\\

Keeping in mind a possible uncertainty in flux calibration between our work and the previously reported measurements, and assuming that our zeropoints are consistent in all of our fields, %
we will measure the [\ion{O}{iii}] luminosity at 6.8 pc ($L_{\rm OIII,6.8})$, or their upper limit, for all our SSSs and SNRs and compare them with those resulting from {\sc cloudy} models integrated to the same radius.
\\*

\begin{figure} 
\includegraphics[width = \columnwidth]{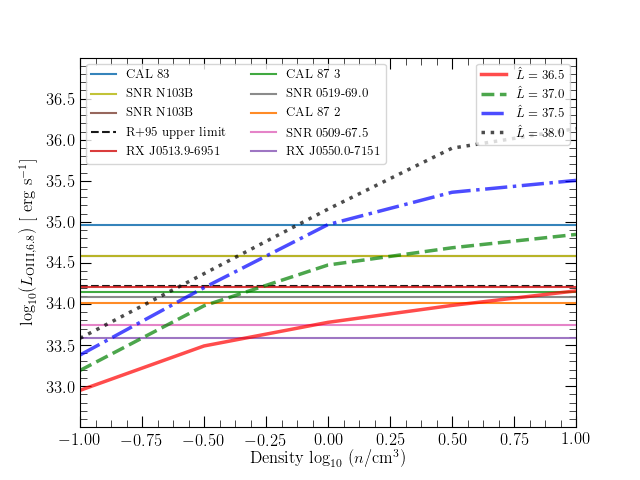}
\caption[ {[}\ion{O}{iii}{]} luminosity at 6.8 pc for all the fields + {\sc cloudy} models]{Logarithm of the [\ion{O}{iii}] luminosity, or its upper limit,
enclosed within 6.8 pc of the central source ($L_{\rm OIII,6.8}$) for all the fields (see Table~\ref{tab:final}), together with the luminosity predicted by four {\sc cloudy} models with $T_5=5$, four luminosities for the central source  $\hat{L}=\{36.5,37.0,37.5,38.0\}$ (corresponding to $\log{(L/L_{\sun})}=$ 2.92, 3.42, 3.92, and 4.42), and nebular densities in the range of $-1.0 \leq \hat{n} \leq 1.0$.
The upper limit to the luminosity of the nebulae undetected by \citet{remillard}, $\hat{L}_{\text{O}}=34.3$, is also plotted.}
\label{fig:Lupperlimits} 
\end{figure}

Figure~\ref{fig:Lupperlimits} presents the results of this exercise.
In order to simplify the analysis, and due to its smaller impact on the results, we use only one radiation temperature for the central source, $T_5=5$. We will allow four values for the luminosity, $\hat{L}=$ 36.5, 37.0, 37.5, and 38.0, which correspond to $\log{(L/L_{\sun})}=$ 2.92, 3.42, 3.92, and 4.42, and allow the more critical parameter, the density of the nebula, to run between 0.1 and 10 cm$^{-3}$.
\\*

A look at the figure, and the summary of results in Table~\ref{tab:final}, confirms that our measurements of CAL 83 agree with those of previous studies.
Our {\sc cloudy} limits for central source luminosity $37.0 \lesssim \hat{L} \lesssim 37.5$ ($3.42 \lesssim \log{L/L_{\sun}} \lesssim 3.92$), leave a range of possible ISM densities for the nebula ($ 1.0 \lesssim n \lesssim 10$ cm$^{-3}$), which is consistent with the range quoted by \citet{remillard} and \citet{gruyters} for their measurements within 6.8 pc ($ 1.0 \lesssim n \lesssim 10$ cm$^{-3}$).
For SNR N103B we do measure some emission, but we have already analyzed the case and concluded that it is not emission associated with the kind of [\ion{O}{iii}]-emitting  ionized region for which we are searching.
For all of the other fields, we have nominal detections but, given the additional uncertainty in our calibration, we interpret these measurements as constraints on any extended nebular emission, all of which are more stringent than those of \citet{remillard}.
As a result, only the assumption of extremely subluminous central sources, with time-averaged luminosities over their recent accretion history of $\hat{L}=36.5$ or less, allows for densities of the surrounding nebulae comparable to that of CAL 83.
%
Alternatively, for central source luminosities persistently comparable to that which is presently observed for these sources ($\hat{L}\gtrsim37$), the required nebular densities are much lower, typical of either lower-density warm ISM or hot phase ISM.
Factoring in the $\sim$45 per cent difference with the flux of \citet{remillard} does not substantially change the previous conclusion.
Our upper limits in this case remain more stringent limits than those of \citet{remillard}.\\

\begin{table}

\caption{[\ion{O}{iii}] luminosity at 6.8 pc}
\begin{threeparttable}
\begin{tabular}{ l  c l}
\toprule\midrule
Object & $L_{\rm OIII,6.8}$\tnote{1} & Remarks\\ \midrule

CAL 83 & 34.96 & real nebula\\
SNR N103B & 34.58 & not a nebula\\
R+95 upper limit & 34.22 & upper limit\\\
RX J0513.9-6951 & 34.20 & nominal detection/upper limit\\
CAL 87 3 & 34.14 & nominal detection/upper limit\\
SNR 0519-69.0 & 34.08 & nominal detection/upper limit\\
CAL 87 2 & 34.01 & nominal detection/upper limit\\
SNR 0509-67.5 & 33.75 & nominal detection/upper limit\\
RX J0550.0-7151 & 33.58 & nominal detection/upper limit\\

\bottomrule\addlinespace[1ex]
\end{tabular}
\begin{tablenotes}\footnotesize
\item[1] Logarithm of the luminosity (erg~s$^{-1}$) at 6.8 pc.
\end{tablenotes}
\end{threeparttable}
\label{tab:final}
\end{table}

\begin{figure} 
\includegraphics[width = \columnwidth]{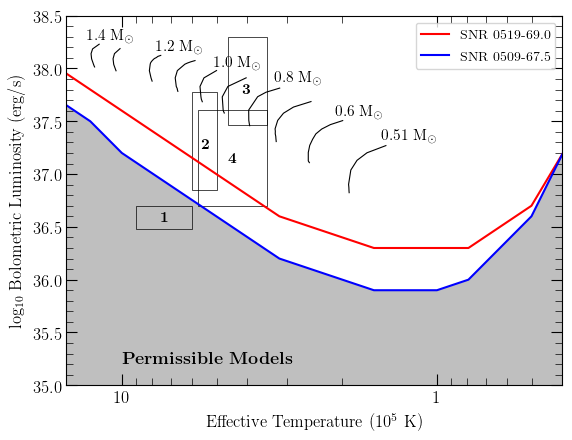}
\caption{Measurements of the luminosity of SNR 0509-67.5 (red) and SNR 0519-69.0 (blue), for a given $T$ (SNR N103B is excluded due to the uncertainty of the source of emission). Black curves are the accreting nuclear-burning WDs models from \citet{wolf2013}, while in black boxes are the estimated ranges of bolometric luminosity and effective temperature of the known close binary SSSs in the LMC: 1. CAL 87; 2. 1E 0035.4-7230; 3. RX J0513.9-6951; and 4. CAL 83~\citep{starrfield,greiner}. The areas below each of the upper limits corresponds to the location of the permissible models for each source.}
\label{fig:HRupperlimits} 
\end{figure}

\begin{table}
\caption{Derived ISM densities around SSSs from the {\sevensize SB} profiles.}
\begin{threeparttable}
\begin{tabular}{ l c c }
\toprule\midrule
Object\tnote{1} &         & $\log_{10} n$ (cm$^{-3}$)\\ \midrule
CAL 83 &   &0.1 \\
CAL 87\tnote{2} &        &< -0.1 \\
RX J0513.9-6951 &        & <-0.6 \\
\bottomrule\addlinespace[1ex]
\end{tabular}
\begin{tablenotes}\footnotesize
\item[1] No $L_{\text{bol}}$ for RX J0550.0-7151.
\item[2] Both epochs.
\end{tablenotes}
\end{threeparttable}
\label{tab:denupper}
\end{table}

In Figure~\ref{fig:HRupperlimits}, we similarly show our results for SNR 0509-67.5 and SNR 0519-69.0. Our limits on the progenitors of these SNe are obtained from searching for the minimum luminosity, at any distance larger than four and smaller than twenty parsecs from the source, of the model (in the bolometric luminosity-effective temperature grid) whose surface brightness profile exceeds the value of the observed surface brightness (we use a smoothed version of the latter to reduce the impact of the noise).
For comparison, in the Hertzsprung-Russell diagram we have also plotted the accreting nuclear-burning WD models of~\citet{wolf2013} (black curves), alongside reliable ranges of the luminosity and temperature of the confirmed
close-binary Magellanic SSS:
1. CAL 87; 2. 1E 0035.4-7230; 3. RX J0513.9-6951; and 4. CAL 83, in black boxes~\citep{starrfield,greiner}. 
It is easy to see the only source that lies in the permissible region of SNR 0509-67.5 is CAL 87, and this source is understood be observed edge-on, and may be more more intrinsically luminous~\citep{starrfield}. For SNR 0519-69.0, the case is different given that its upper limit is consistent with CAL 83 and 1E 0035.4-7230.\\ 

It is notable that the obtained upper limits are less stringent than the ones from~\citet{kuuttila2019} and~\citet{woods2018} (though independent), likely due to the aforementioned contamination in the difference images. The [\ion{O}{iii}]-derived upper limits for the progenitor of SNR 0509-67.5 are comparable to that obtained by~\citet{graurwoods} for the SN 2014J, using a similar approach. 
\\*

Lastly, for CAL 83, CAL 87 and RX J0513.0-6951, the SSSs in our sample with known bolometric luminosities and effective temperatures, we have determined the model with the largest (surrounding) density whose {\sevensize SB} profile exceeds the value of our smoothed observed {\sevensize SB},  at any distance between one and twenty parsecs, assuming their presently-observed luminosities and effective temperatures have remained persistent over the last $\sim10^4$ years, i.e., the $\rm{O}^{2+}$ recombination time in 0.1--1 $\rm{cm}^{-3}$ gas. The results are shown in Table~\ref{tab:denupper}; it is readily apparent that if these sources are indeed persistent over long time scales, the ambient surrounding ISM densities inferred for CAL 87 and RX J0513.9-6951 are surely lower than for CAL 83, potentially indicating they are located well above the midplane of the LMC or in a region of hot ISM.

\section{Discussion}

Our results are consistent with the earlier work of \citet{remillard}. The only detected [\ion{O}{iii}] emission nebula around a SSS is that of CAL 83.
Although we manage to establish more sensitive upper limits, no [\ion{O}{iii}] ionized regions are detected with confidence around the other SSSs or SNRs.
The exception, where some emission is measured, is SNR N103B.
But, as discussed above, the signal in this case does not correspond to a diffuse, fossil [\ion{O}{iii}] ionized nebula associated with the central source but to compact emission knots inside the expanding shock of the SN and a to a nearby superbubble \citep{williams2014}. 
\\*

One possible interpretation of the absence of other detected nebulae is that most SSSs undergo luminosity variations on much longer timescales than have presently been observed ($\sim$ decades), in which they are primarily in the X-ray off state.
Any variability that is shorter than a tenth of recombination timescale of hydrogen (recall Equation~\ref{eq:rec}),
should not strongly affect the ionization nebula {\sevensize SB} profiles~\citep[][{though note that they do not investigate $[\ion{O}{iii}]$ emission, but consider a constant-temperature nebula of H and He only}]{chiang}.
Given their presently observed duty cycles, the X-ray on (optically low) states of CAL 83 and RX J0513.9-6951 are $\sim 200$ and $\sim40$ days, while the X-rays off states (optically high) are $\sim 250$~ and $\sim 140$ days, respectively~\citep{rajoelimanana}. In these cases, then, detectable ionization nebulae are expected for both if the ISM density is sufficiently high.
Therefore, either the time-averaged luminosity of RX J0513.9-6951 (and all of the other SSSs in the sample but CAL 83) is lower than 
the present value, i.e., there is a yet-unseen long term variability or long \enquote*{duty cycles}, or e.g., the ISM surrounding them is much less dense than in the vicinity of CAL 83. The \enquote*{duty cycle} argument is also valid to explain the lack of emission in the SNR case, but notable for SNRs the densities are known.  
\\*

Another possibility is the action of an optically thick wind reacting to accretion rates above the stable-burning region and obscuring the X-ray source~\citep{hachisu99a}, however this too should be observable, as the fast wind would excavate a cavity in the surrounding ISM and produce a dense expanding shell~\citep[see discussion in][]{woods2016}.
We are then led to conclude that either the ISM densities surrounding CAL 87, RX J0513.9-6951 and RX J0550.0-7151 are about $\sim 0.1 \text{ cm}^{-3}$, or less, much lower than that of CAL 83, or that there is some other evolutionary process which obscures the soft X-ray emission of the sources and prevents the formation of the nebulae \citep[see also][for further discussion]{kuuttila2019}.
\\*

The idea that the detection of ionized nebulae would help us to find an obscured SSS that cannot be detected via X-rays \citep{remillard}, does not consider the possibility that the ionizing radiation could not reach the ISM around the source.
The issue has been considered in studies trying to understand the discrepancy between the observed and expected number of SSSs in galaxies.
We note \citet{gilfanov2010}, who focus on elliptical galaxies, and \citet{distefano2010a}, who focus on spiral ones.
\citet{nielsen2013} proposed that the circumbinary stellar medium (CSM) surrounding a SSS should be able to 
obscure most of the radiation by the action of large mass loss from the system via a dense wind.
This idea was rejected by \citet{nielsen2015}, who used {\sc cloudy} to prove that these hypothetical nebulae would look very different from the expected SSS nebulae \citep{rappaport}, with higher \ion{He}{ii}/H$\beta$ and lower [\ion{O}{iii}]/H$\beta$ ratios, while requiring mass loss rates comparable to the stable-burning accretion rate ($\sim 10^{-7} - 10^{-6} M_{\sun}$) for a typical SSS system\footnotemark\footnotetext{ 
$k T_{\text{eff}} = 50 $eV, $L_{\text{bol}} = 10^{38}$ erg s$^{-1}$, inner/outer radial distances to the compact source $r_{\text{inn}} = 1$ AU, 
$r_{\text{out}} = 10$ AU, and wind speed $u_{\text{w}} = 10$ km/s.}. The amounts of CSM material implied by this scenario would imprint features in the early spectrum of a SN Ia, which are very seldom seen \citep[see][for the case of SN 2006X]{patat2007}. In particular, the lack of X-ray and radio detections from early observations of SNe Ia strongly constrains the mass loss rate from their progenitor systems, typically excluding symbiotic stars and even fast optically-thick winds \citep[e.g.,][]{Chomiuk2016,Lundqvist2020}.\\

With respect to the SNRs, our results are consistent with those of \citet{kuuttila2019}, although they use a different approach.
They measured the surface brightness upper limits from the spectrum of the three SNRs, SNR 0509-67.5, SNR 0519-69.0 and SNR 0509-68.7,  using the \ion{He}{ii}$\lambda$4686 {\AA}  emission line from spectra taken at $\sim 4-5$ pc from the central sources.
They did not find emission in any of the SNRs, and from the upper limit of \ion{He}{ii} emission of SNR 0519-69.0, and the constraints on the companion luminosity given by \citet{edwards}, they
ruled out any possibility of a single degenerate progenitor, including SSSs.
For SNR 0509-67.5, our results also agree with \citet{woods2018} who have independently excluded the possibility of a SSS progenitor given the high content of neutral hydrogen in the surrounding environment.
Note that \cite{kuuttila2019} also discussed the possibility  of constraining SNR progenitor luminosities by setting upper limits to the [\ion{O}{iii}] emission,  in particular noting the [\ion{O}{iii}]$\lambda 5007$ line visible in the spectrum of SNR  0519-69.0 as an example.
However, they noted the forward shock itself would contribute some UV-soft and X-ray radiation and this makes it necessary to model the whole SNR spectrum in order to separate a hypothetical relic nebulae in the immediate vicinity of the shock from the diffuse emission resulting from the ejecta.
We avoid this ambiguity 1) by observing well beyond the shock front, and 2) because we claim only upper limits on a pre-existing fossil nebula (whereas a detection of such a nebula would require much more sophiscated modeling in order to distangle the shock-induced emission from the fossil nebula).

\section{Conclusions}

Although we are within $3\sigma$ of the results in the literature, the uncertainty in flux calibration prevents us from making a firm conclusion regarding the flux of any putative very low surface brightness nebulae for the sources in our sample (other than CAL 83).
Practically speaking, going deeper using our present method would require background determination for our image subtraction to be accurate to within a fraction of an ADU across the whole field of view, and the difference between our measurements and those of \citet{remillard} amounts to just $ \lesssim 1 $ ADU per pixel (2$\sigma$ at 6.8 pc, exceeding literature value at 22.7 pc), without considering a better replacement of the masked values, notably near the source (see Figure~\ref{fig:fluxcal83}). 
Particularly, we are concerned about the statistics provided by {\sc hotpants} when subtracting images taken with different filters, since even in the optimal cases we tried, such as the subtraction between $V-V$ and O$-$O filters of consecutive images, the results were puzzling to us.
\\*

Qualitatively, however, our results are fully aligned with those of previous works.
Neither the SSSs (aside from CAL 83) nor the SNRs yielded a new detection of an  [\ion{O}{iii}]-luminous ionized nebula.
In addition, we have obtained the following new results: 

\begin{enumerate}

\item We have presented {\sevensize SB} profiles around seven hot and luminous sources in the LMC, four SSSs and three SNRs.
Notably, although CAL 83 was first studied in detail in 1995, our {\sevensize SB} profile is to our knowledge the first published. 

 \item The [\ion{O}{iii}] flux measurement of the CAL 83 nebula at 6.8 pc is broadly ($3\sigma$) consistent with those in the literature.
 They reported a 50 per cent increase in luminosity between 6.8 and 22.7 pc and we measure about a 50 per cent increase, approximately, between the same radii, although reduced by a factor of two.
 
\item The shape of the [\ion{O}{iii}] SB profile is only crudely approximated by constant density {\sc cloudy} models, consistent with previous studies which have determined that the ISM density in the nebula is not constant. 
The density of the ISM for CAL 83 at 6.8 pc is $\geq 1 \text{ cm}^{-3}$ for a persistent blackbody temperature of $T=5\times 10^{5}$ K and a luminosity of
 $L \leq 10^{37.5}$ erg/s, as observed for the source.
 This is consistent with the calculations by \citet[][$\approx 4-10 \text{ cm}^{-3}$ for the inner nebula]{gruyters} using the values for the density, luminosity and temperature of the
 photoionized models from \citet{rappaport}.
 
 \item The [\ion{O}{iii}] luminosities at 6.8 pc for almost all the fields are below $10^{34.2}$ erg s$^{-1}$, which is about $17$ per cent of our measurement of the luminosity of the CAL 83 nebula.

 \item For the SSSs, the negative detections require either a much lower density ISM compared with that of CAL 83 (for CAL 87, $\lesssim 0.8 \text{ cm}^{-3}$; for RX J0513.9-6951, $\lesssim 0.25 \text{ cm}^{-3}$), or a very low ionizing luminosity averaged over the last $\sim$ recombination timescale compared to their currently observed state. 
This is the first time such non-detections have been used to place a strong upper bound on the ISM density in the vicinity of CAL 87 and RX J0513.9-6951, providing a unique probe of the warm diffuse ionized medium in the LMC under the assumption that the presently observed SSS spectra are typical of at least the last several thousand years.

\item For the SNRs, the expanding remnant itself provides an independent probe of the surrounding ISM density; for this reason, we may unambiguously exclude any persistently hot and luminous progenitor within the last $\sim$10,000 years. This result is consistent with other independent means of constraining their progenitors, 
\citep[e.g.,][]{woods2018,kuuttila2019,graurwoods}.
\end{enumerate}

\section*{Acknowledgements}
We thank the referee for their kind and insightful comments, which greatly improved the manuscript. This paper includes data gathered with the 6.5 m Magellan Telescopes located at Las Campanas Observatory, Chile. This research has made use of NASA's Astrophysics Data System. Support for DNF and AC was provided by ANID, through the Millennium Science Initiative grant ICN12\_009 (MAS), and from grant Basal CATA PFB 06/09. TEW acknowledges support from the NRC-Canada Plaskett fellowship.

\section*{Data Availability}
The data underlying this article will be shared on reasonable request to the corresponding author.




\bibliographystyle{mnras}
\bibliography{bibliografia} 


\bsp	
\label{lastpage}
\end{document}